\documentclass[aps,twocolumn,preprintnumbers]{revtex4}

\usepackage{etoolbox} 
\usepackage{soul}
\usepackage{graphicx}  
\usepackage{subfigure}
\usepackage{multirow}
\usepackage{tikz,lipsum,lmodern}
\usepackage{orcidlink} 
\usepackage{amsmath}
\usepackage{amssymb}
\usepackage{graphicx}
\usepackage{siunitx}
\usepackage{indentfirst}
\usepackage{orcidlink} 
\usepackage{mhchem}
\usepackage{xcolor}
\usepackage{graphicx}  
\usepackage{orcidlink} 
\usepackage{subfigure}
\usepackage{multirow}
\usepackage{mhchem}
\usepackage{natbib}
\usepackage{fancyhdr}
\usepackage{longtable}
\usepackage{parskip}
\usepackage[T1]{fontenc}
\usepackage{dcolumn}   
\usepackage{bm}        
\usepackage{amsfonts}  
\usepackage{amsmath}   
\usepackage{amssymb}   

\newcommand{\pwisein}{\left\{ \begin{array}{ll}}
\newcommand{\pwiseout}{\end{array}\right.}

\begin{document}

\title{Prediction and Experimental Verification of Electrolyte Solvation Structure from an OMol25-Trained Interatomic Potential}



\author{Nitesh Kumar$^{a,b}$\orcidlink{0000-0003-3322-8450}}

\author{Jianwei Lai$^{a,c,g}$\orcidlink{0000-0001-5841-1134}}
\author{Casey S. Mezerkor$^{a,d,f}$\orcidlink{0009-0002-7672-5027}}
\author{Jiaqi Wang$^{a,d}$}
\author{Kamila M. Wiaderek$^{a,e}$\orcidlink{0000-0002-0051-3661}}
\author{J. David Bazak$^{a,f}$\orcidlink{0000-0002-4599-3208}}
\author{Samuel M. Blau$^{a,g,h}$\orcidlink{0000-0003-3132-3032}}
\email{smblau@lbl.gov}
\author{Ethan J.\ Crumlin$^{a,c,i}$\orcidlink{0000-0003-3132-190X}}
\email{ejcrumlin@lbl.gov}

\affiliation{$^{a}$ Energy Storage Research Alliance, Argonne National Laboratory, Lemont, Illinois 60439, USA}

\affiliation{$^{b}$ Materials Sciences Division, Lawrence Berkeley National Laboratory, Berkeley, California, USA}

\affiliation{$^{c}$ Chemical Sciences Division, Lawrence Berkeley National Laboratory, Berkeley, California, USA}

\affiliation{$^{d}$ Department of Chemistry and Biochemistry and the Oregon Center for Electrochemistry, University of Oregon, Eugene, Oregon 97403, USA}

\affiliation{$^{e}$ X-ray Science Division, Argonne National Laboratory, Lemont, Illinois 60439, USA}

\affiliation{$^{f}$ Physical and Computational Sciences Directorate, Pacific Northwest National Laboratory, Richland, Washington 99354, USA}

\affiliation{$^{g}$ Energy Technologies Area, Lawrence Berkeley National Laboratory, Berkeley, California, USA}

\affiliation{$^{h}$ Bakar Institute of Digital Materials for the Planet, University of California, Berkeley, California, USA}

\affiliation{$^{i}$ Advanced Light Source, Lawrence Berkeley National Laboratory, Berkeley, California, USA}

\date{\today}

\begin{abstract}  

A molecular-level understanding of electrolyte solvation structure and ion–ion correlations is critical to developing next-generation battery chemistries. Atomistic simulation capabilities with sufficient accuracy, speed, and transferability to deliver reliable structural insights while avoiding arduous system-specific re-parameterization are thus highly desirable. 
Machine learning interatomic potentials (MLIPs) trained on large, chemically diverse datasets are revolutionizing computational chemistry, enabling molecular dynamics simulations of battery electrolytes with near-DFT accuracy over 10,000$\times$ faster than DFT. While previous MLIP training datasets with suitable elemental coverage for electrolytes have been based on inorganic materials, the Open Molecules 2025 (OMol25) dataset provides large-scale molecular DFT MLIP training data with broad elemental coverage and specifically samples tens of millions of electrolyte configurations. 
Here, we integrate computational modeling with experimental validation to systematically assess the ability of large-scale MLIPs pre-trained on materials data or on OMol25 to accurately resolve nanoscale structural organization and ion-solvation characteristics in Na-ion battery electrolytes across diverse physicochemical conditions and compositional regimes. We find that the OMol25-trained Universal Model of Atoms (UMA-OMol) predicts experimentally measured densities and X-ray structure factors in substantially better agreement compared to state-of-the-art models trained only on inorganic materials data. Using UMA-OMol, we further analyze systematic trends in solvation structure as a function of cation identity, anion chemistry, salt concentration, and solvent topology. We observe that increasing system temperature amplifies the heterogeneity within the solvation environment, perturbing cation–solvent interactions and promoting the formation of contact ion pairs (CIPs). Moreover, subtle variations in the solvent topology of glyme-based electrolytes cause pronounced changes in ion-correlations and solvation structure. 
The experimental agreement and microscopic insights shown here position OMol25-trained MLIPs as a practical route to predictive, high-throughput electrolyte simulations beyond the limits of classical force fields and direct DFT molecular dynamics, serving as a powerful tool for accelerating the design of next-generation Na-ion battery electrolytes and beyond.

\textbf{Keywords:} \textit{sodium-ion battery, machine learning interatomic potentials, MLIPs, OMol25, UMA-OMol, molecular dynamics, electrolyte structure}

\end{abstract}

\pacs{47.15.-x}

\maketitle

Accurately capturing microscopic solvation structure is crucial to predicting macroscopic properties in complex liquid mixtures like battery electrolytes. Predictive atomistic simulations of electrolyte solvation structure could dramatically accelerate electrolyte discovery and additive engineering via high-throughput \textit{in-silico} screening or by elucidating mechanistic insight that enables rational chemical design.\cite{schran2021machine,kumar2023adsorbate,zhu2024differentiable,magduau2025predictive} 

Over the past several decades, the pursuit of chemically accurate molecular dynamics (MD) for complex, multicomponent systems has been constrained by a fundamental trade-off between accuracy and scale.\cite{mouvet2022recent} \textit{Ab-initio} MD captures the delicate interplay of covalent and non-covalent interactions with quantum-mechanical rigor, but its prohibitive computational cost limits accessible length and time scales to regimes far from those relevant to mesoscale structure and transport.\cite{kumar2021essential} Classical and polarizable force fields, while computationally efficient, demand labor-intensive parameterization and are often anchored to a narrow domain of validity, relying on empirical adjustments that cannot fully capture the many-body effects, electronic polarization, and subtle free-energy landscapes of chemically heterogeneous environments.\cite{wildman2016general} 

Recent advances in machine learning interatomic potentials (MLIPs) trained on density functional theory (DFT) data offer a compelling path forward: near-\textit{ab-initio} accuracy at a fraction of the computational cost, with the flexibility to study both covalent and non-covalent structure in response to the local chemical environment.\cite{jacobs2025practical,batatia2022mace} Yet, as with any ML model, their fidelity is ultimately bounded by the size, chemical diversity, and consistency of the training data. While MLIPs trained on small, system-specific datasets can yield novel insight into individual systems of interest,\cite{gartner2020signatures,kumar2025structure,wang2024accelerating} the time and resources required to generate DFT training data and train models is often onerous, and the resulting trained models are not transferable. In contrast, MLIPs pre-trained on very large, diverse datasets seek to provide sufficiently accurate predictions across chemical space to obviate the need for system-specific data generation and model training.\cite{batatia2025foundation,deng2023chgnet} However, widely used large-scale molecular DFT datasets like SPICE/2\cite{eastman2022spice,spice2zenodo} and AIMNet2 have little to no coverage of critical metals for battery electrolytes (e.g. Li, Na, Zn) while electrolyte-specific works like BAMBOO\cite{gong2025bamboo} and Wang et al.\cite{Wang2025} focus almost exclusively on Li-ion systems. Thus, the only large-scale pre-trained MLIPs with sufficient elemental coverage for diverse liquid electrolytes have paradoxically been those trained on crystalline material planewave DFT training data from the Materials Project\cite{jain2013materialsproject} (MPtrj),\cite{deng2023chgnet} Alexandria\cite{schmidt2024alexandria} and/or Open Materials 2024 (OMat24)\cite{barroso2024open,horton2025accelerated}. However, the recently released Open Molecules 2025 (OMol25) now provides a large-scale, high-accuracy molecular DFT dataset for training MLIPs with extensive elemental coverage and directly targets electrolytes as one of its primary areas of focus.\cite{ju2025application}

In this study, we evaluate the predictive capability of large-scale, pre-trained MLIPs for describing the macroscopic ensemble averaged properties and the microscopic interatomic interaction patterns that govern the structure and dynamics of next-generation Na-ion battery electrolytes. While Li-ion electrolytes have dominated consumer electronics for decades, Na-ion batteries present major advantages in cost and material availability, motivating computational discovery.\cite{hwang2017sodium,yabuuchi2014research} We investigate pre-trained MLIP performance for simulating \ce{NaPF6} (sodium hexafluorophosphate), NaOTf (sodium triflate) or NaTFSI (sodium bis(trifluoromethanesulfonimide)) in ether and carbonate solvent matrices and under different system conditions such as temperature or concentration. Further, only large-scale MLIPs pre-trained on either inorganic materials DFT data or on a sufficiently diverse molecular dataset like OMol25 are viable for simulating such systems.

We benchmark three large-scale, pre-trained MLIPs, two trained on the inorganic materials OMat24 dataset -- Orb-OMat\cite{neumann2024orb} and SevenNet-OMat\cite{park_scalable_2024} -- and one trained on OMol25 -- UMA-OMol\cite{levine2025openmol,wood2025family} -- against experimental Na-ion electrolyte density and X-ray structure factor measurements, establishing that UMA-OMol most accurately reproduces these macroscopic properties. We then examine the performance of UMA-OMol across varying battery operating temperatures, where Na-ion systems are compared to well-understood Li-ion systems as a baseline, followed by a systematic study of anion identity to capture the correct hierarchy of anion–cation interaction strengths (via coordination number) and their influence on solvation structure. We further investigate the concentration dependence of \ce{NaPF6} electrolytes, quantifying the shift from free-ion populations to contact and solvent-separated ion pairs and linking these structural changes to transport-relevant trends observed in experiment. This is followed by an investigation of solvent effects \cite{kumar2024role} on ion pairing and aggregation across six different solvent candidates for Na-ion batteries, and subsequent analysis of atomic-scale structuring at the graphite interface.

Our results show that the OMol25-trained MLIP UMA-OMol is an efficient surrogate for first principles simulations of battery electrolytes and offers broad promise for predictive modeling of complex electrochemical systems, providing chemically transferable, \textit{ab-initio}-level predictions of nanoscale electrolyte behavior at orders of magnitude reduced computational cost versus DFT-based molecular dynamics. Further, UMA-OMol shows performance superior to models trained on inorganic materials DFT datasets that did not include liquid-phase, electrolyte-specific structural sampling. Overall, we view OMol25-trained models as powerful first line tools for modeling battery electrolyte systems, to be complemented as needed by targeted electronic structure calculations and experimental data as model architectures continue to improve.

\begin{figure*}[ht!]
\centering
  \includegraphics[height=16cm]{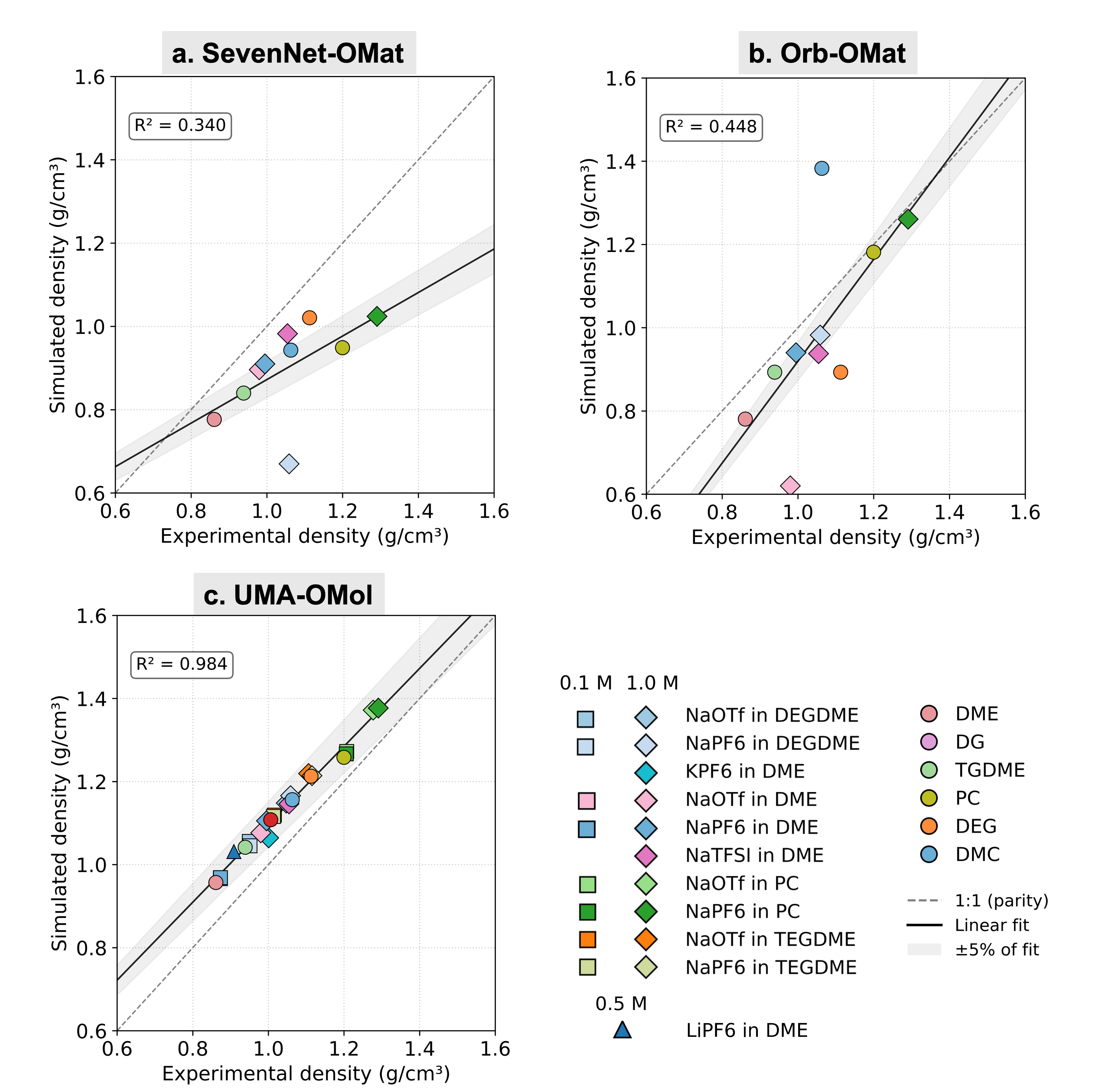}
  \caption{Pre-trained MLIP predicted liquid density comparisons for different organic and sodium electrolyte systems. Experimental densities are compared with simulated densities from two inorganic materials-trained models (a) SevenNet-OMat (b) Orb-OMat and one OMol25-trained model (c) UMA-OMol across battery electrolytes that vary in salt-type, salt-concentration, and solvent-type. Here, \ce{OTf}, \ce{TFSI}, and \ce{PF6} correspond to triflate, bis(trifluoromethanesulfonimide), and hexafluorophosphate anions respectively. DME, DEGDME, PC, TEGDME, DEG, and DMC correspond to dimethoxyethane, diethylene glycol dimethyl ether (diglyme), propylene carbonate, tetraethylene glycol dimethyl ether, diethylene glycol, and dimethyl carbonate respectively. The dashed line represents the ideal parity relation between simulation and experiment. The solid line denotes the least squares linear regression fit to the simulated versus experimental densities. The shaded region indicates a ±5\% envelope around the fitted regression line, highlighting the acceptable deviation window relative to the model trend.}   
  \label{fgr:omol_dens}
\end{figure*}

\begin{figure*}[ht!]
\centering
  \includegraphics[height=9.5cm]{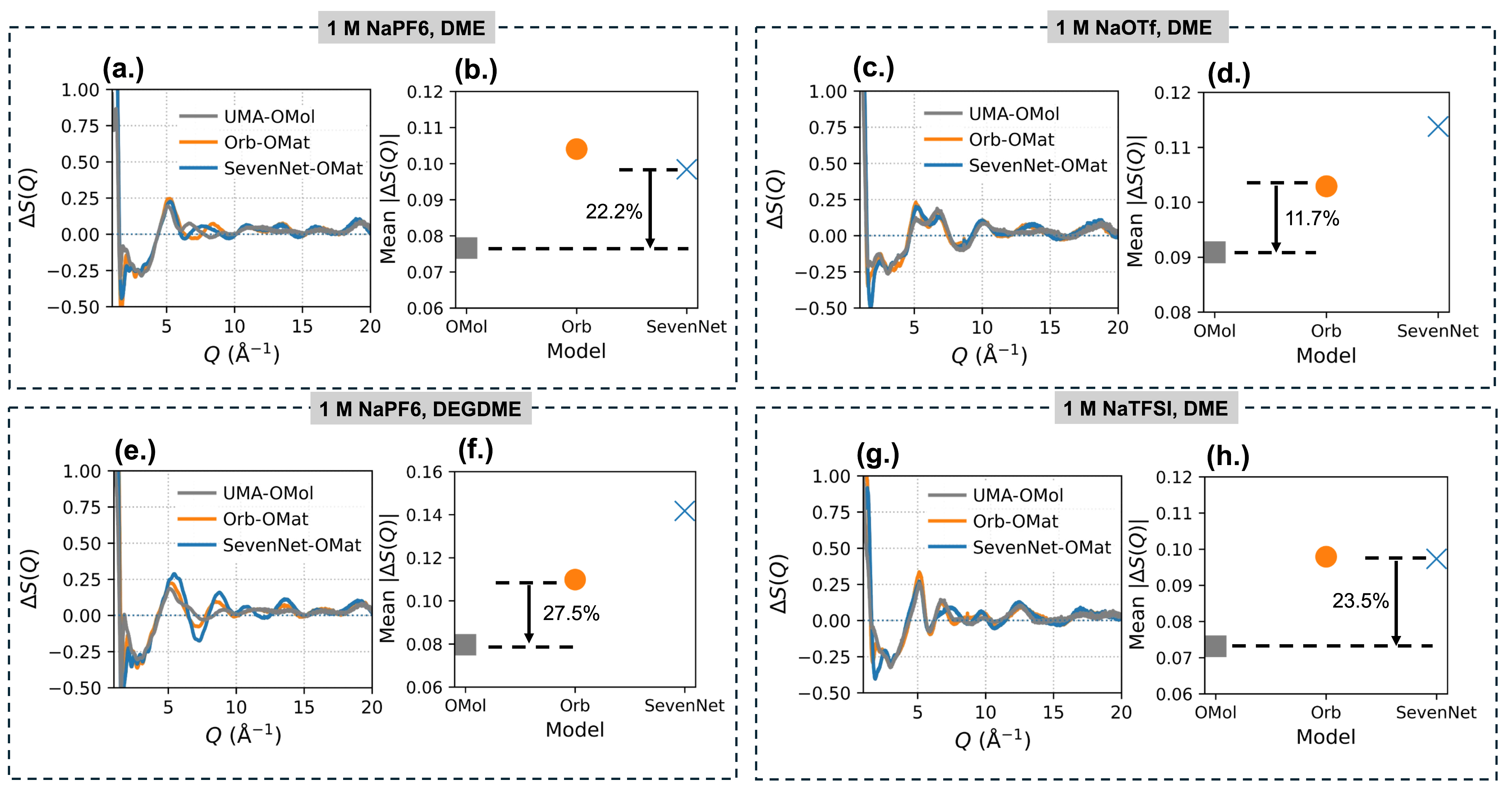}
  \caption{Residual X-ray structure factors and mean deviations for 1 M sodium salt glyme electrolytes. Panels (a), (c), (e), and (g) show the difference between simulated and experimental structure factors, i.e. the residual 
\(\Delta S(Q) = S_{\mathrm{sim}}(Q) - S_{\mathrm{exp}}(Q)\) 
for NaPF\(_6\) in dimethoxyethane (DME), NaOTf in DME, NaPF\(_6\) in diglyme (DEGDME), and NaTFSI in DME, respectively, comparing UMA-OMol (gray), Orb-OMat (orange), and SevenNet-OMat (blue) over \(Q = 1\) to \(20\ \text{\AA}^{-1}\).  
Panels (b), (d), (f), and (h) present the corresponding mean absolute deviations \(\langle |\Delta S(Q)| \rangle\), showing that UMA-OMol systematically yields the smallest residuals, with Orb-OMat often intermediate and SevenNet-OMat often largest.}   
  \label{fgr:sq_diff}
\end{figure*}

\paragraph{\ul{Inorganic Materials vs. OMol25 Pre-trained MLIPs}}

It is widely accepted that the fundamental thermodynamic property of a battery electrolyte that directly influences the structure, dynamics, and thermodynamic behavior of the system is the system density.\cite{hansen2013theory_simple_liquids,han2017salient} Accurately capturing experimental densities is essential for meaningful comparison and reliable prediction of material behavior under realistic conditions. Accordingly, we measured the densities of pure solvents and Na-ion battery electrolytes across different salt types, concentrations, and solvent environments, as well as one Li-ion and one K-ion system, and compared them with densities predicted by NPT molecular dynamics driven by UMA-OMol, Orb-OMat, and SevenNet-OMat in Figure~\ref{fgr:omol_dens}, and the respective numerical values are provided in Table S1. Simulations performed using the SevenNet-OMat and Orb-OMat large-scale MLIPs trained on inorganic materials datasets MPtrj,\cite{deng2023chgnet} Alexandria,\cite{schmidt2024alexandria} and OMat24\cite{barroso_luque2024omat24}) exhibit substantial deviations from experimental densities (see Figure \ref{fgr:omol_dens}a and b). For SevenNet-OMat, the simulated densities show systematic underestimation and significant scatter ($R^2 \approx 0.34$), while Orb-OMat exhibits only slightly better agreement with experiment ($R^2 \approx 0.45$). Further, for both models we observed several MD simulations become unstable, causing an unphysical breaking of the simulation box and precluding the trajectory from ever reaching equilibrium. We hypothesize that these deficiencies stem from the absence of liquid-phase configurations and solvation environments in the inorganic training datasets, limiting the model’s ability to capture subtle intermolecular and ion–solvent correlations.

In contrast, the parity plot between experimental and UMA-OMol predicted densities shows excellent agreement across linear and cyclic ethers, carbonates, and their 0.1 M and 1.0 M sodium salt solutions (\ce{NaPF6}, \ce{NaOTf}, \ce{NaTFSI}, etc.), yielding an overall coefficient of determination \(R^{2} = 0.98\). We do observe a systematic shift in the UMA-OMol parity toward slightly higher values across all electrolyte systems. On average, the simulated densities are about 8.5 \% higher than experiment, which we attribute to the combination of known overbinding of many-body dispersion by the $\omega$B97M-V\cite{mardirossian2016wb97mv,furness2022r2scan_rvv10,arulmozhiraja2020rvv10_overbinding} density functional used to construct the OMol25 dataset and the absence of explicit long range interactions in UMA-OMol. We observed a similar behavior for density variations with temperature, as shown in Figure S1, which compares experimental and simulated densities for 0.5 M \ce{NaPF6} and \ce{LiPF6} electrolytes in DME. Despite the systematic shift, both systems exhibit the expected linear thermal expansion, with density decreasing as temperature increases, and our simulations reproduce this trend with high fidelity. On average, the simulated densities are about 10 \% higher than experiment for both electrolytes at different temperatures, reflecting a modest overestimation of solvent packing and ion associated partial molar volumes. The narrow confidence band around the parity line further demonstrates consistent transferability across chemically distinct electrolytes, highlighting that UMA-OMol accurately predicts both neutral and ion-containing systems without any reparameterization. The pronounced improvement from inorganic materials trained models to UMA-OMol underscores the importance of explicitly incorporating liquid state chemistries in the MLIP training dataset and the higher fidelity density functional accessible in a molecular context.

We further assessed the molecular structuring predicted by UMA-OMol and other models against experimental X-ray structure factors \(S(Q)\) for a series of 1 M sodium salt electrolytes (Figure S2-3). The data set comprises NaPF\(_6\), NaOTf, and NaTFSI dissolved in dimethoxyethane (DME), diglyme (DEGDME), and tetraglyme (TEGDME), with each panel reporting one salt solvent combination. To quantitatively measure model accuracy, we evaluated residual structure factors \(\Delta S(Q) = S_{\mathrm{sim}}(Q) - S_{\mathrm{exp}}(Q)\) over \(Q = 1\) to \(20\ \text{\AA}^{-1}\) (Figure~\ref{fgr:sq_diff}). UMA-OMol captures not only the dominant features of \(S(Q)\) but also the high \(Q\) oscillations that arise from bond-stretching, bond-angle fluctuations, and torsional motions of the ether chains, demonstrating an accurate short-range potential energy surface and realistic molecular flexibility. This fidelity is reflected in the mean absolute deviation \(\langle |\Delta S(Q)| \rangle\), which is smallest for UMA-OMol and systematically larger for Orb-OMat and SevenNet-OMat across all electrolytes. Only four of the six systems are shown in Figure~\ref{fgr:sq_diff} because Orb-OMat and SevenNet-OMat MD simulations of NaOTf in DEGDME and NaPF6 in TEGDME were consistently unstable. Quantitatively, UMA-OMol shows a 10–30\% improvement in reproducing molecular structure compared to MLIPs trained on materials data. Noticeable deviations remain in the very low \(Q\) region, where long-range density correlations dominate. These deviations are consistent with the use of a \(6\ \text{\AA}\) cutoff and the absence of explicit long-range electrostatic interactions, which limits the description of correlations at larger length scales. The reduced residuals and well-resolved oscillatory features at intermediate and high \(Q\) indicate that UMA-OMol treats bonded interactions and short-range repulsion more accurately, which in turn leads to enhanced robustness in long molecular dynamics trajectories with minimal bond distortion and few unphysical structural instabilities.

Density and X-ray structure factor results show that UMA-OMol reproduces the structural properties of battery electrolytes more robustly than models trained on inorganic materials data. We now proceed to employ UMA-OMol to interrogate molecular structure across different physicochemical conditions, and variations in anion-type and solvent topology.

\paragraph{\ul{Anion Effects}}

\begin{figure}[ht!]
\centering
  \includegraphics[height=15.0cm]{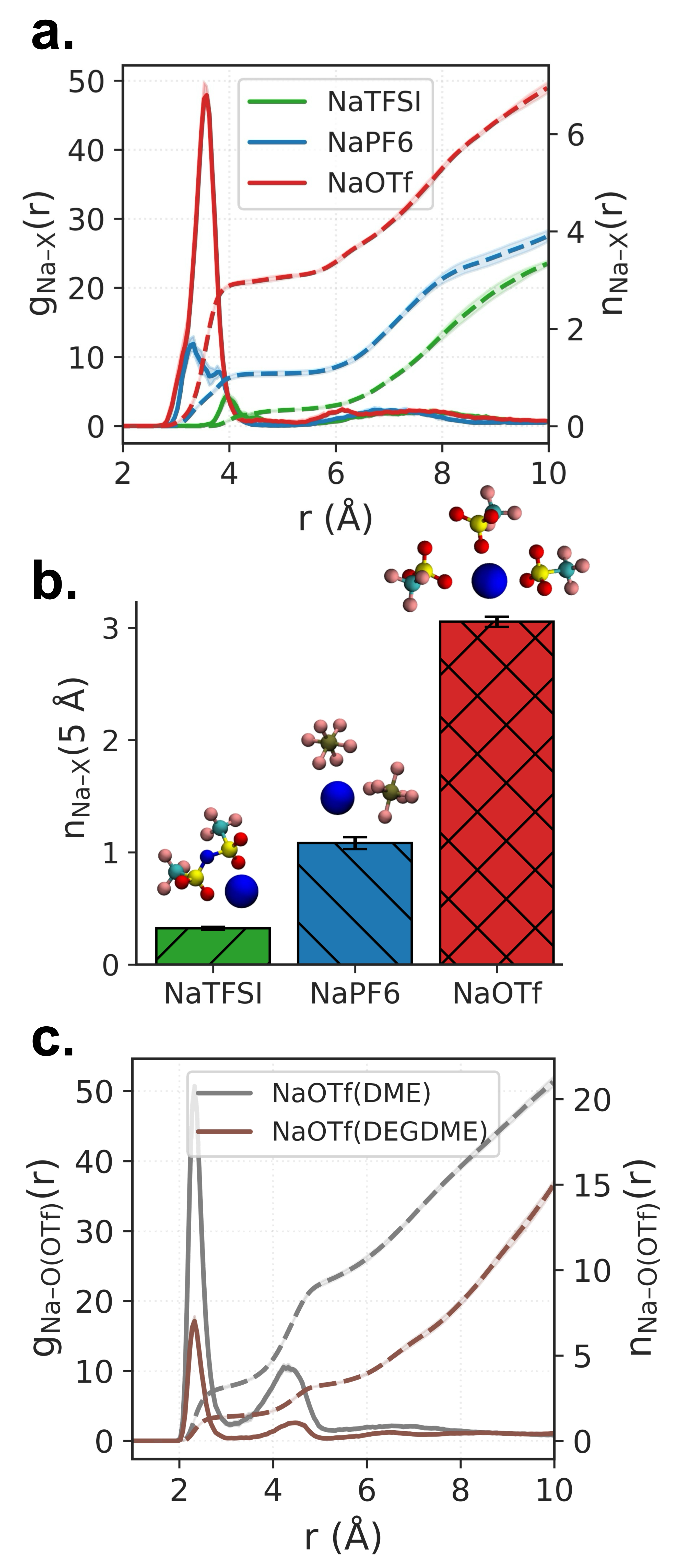}
\caption{Anion-dependent Na$^+$ coordination in DME from UMA-OMol simulations. (a) Radial distribution functions $g_{\mathrm{Na-X}}(r)$ (solid lines, left axis) and running coordination numbers (dashed lines, right axis) $n_{\mathrm{Na-X}}(r)$ for bis(trifluoromethanesulfonyl)imide (TFSI$^-$), hexafluorophosphate (PF$6^-$), and triflate (OTf$^-$), computed relative to N, P and S atoms of TFSI$^-$, PF$6^-$ and OTf$^-$ respectively. (b) Average coordination numbers $n_{\mathrm{Na-X}}(5 Å)$ in primary cation coordination shell highlighting the hierarchy of ion-pairing strength across anions. (c) Radial distribution functions $g_{\mathrm{Na\text{-}O(OTf)}}(r)$ and corresponding running coordination numbers $n_{\mathrm{Na\text{-}O(OTf)}}(r)$ for NaOTf in DME and diglyme (DEGDME) media, demonstrating that solvent identity can modulate ion interaction strength. 
}
\label{fgr:anion}
\end{figure}

We compare Na$^{+}$ association with three common battery anions in glyme electrolytes using radial distribution functions, $g(r)$, and running coordination numbers, $n(r)$, as these interactions directly control electrolyte speciation in Na-ion batteries. Figure \ref{fgr:anion}a shows Na$^{+}$ correlations with TFSI$^{-}$, PF$_6^{-}$, and OTf$^{-}$, referenced to their central N, P, and S atoms, respectively, in 1 M DME electrolytes. The first solvation shell exhibits a clear hierarchy in ion pairing strength, with OTf$^{-}$ displaying a sharp and intense peak at $r \approx 3.4$ Å ($g_{\max} \approx 45$), PF$_6^{-}$ showing a broader and weaker maximum at a similar distance ($g_{\max} \approx 11$), and TFSI$^{-}$ exhibiting only weak structuring ($g_{\max} \approx 4$). Integration of the corresponding coordination numbers to 5 Å (Figure \ref{fgr:anion}b) yields $n = 3.0 \pm 0.1$ for OTf$^{-}$, $1.1 \pm 0.1$ for PF$_6^{-}$, and $0.3 \pm 0.05$ for TFSI$^{-}$, indicating that Na$^{+}$ forms substantially more frequent and experimentally observed stable ion paired configurations with OTf$^{-}$ than with the more weakly coordinating anions, as expected from previous work.\cite{yadav2024direct}

Atom-resolved binding analysis further supports this hierarchy. We computed Na–X radial distribution functions and coordination numbers for the dominant coordinating sites in each anion (X = O in TFSI$^{-}$, F in PF$_6^{-}$, and O in OTf$^{-}$; Figure S4).
Across electrolytes, the contact ion pairing signature is most pronounced for \ce{Na+} binding to O in OTf$^{-}$, which exhibits a very sharp first-shell peak centered near \(\sim 2.3\,\text{\AA}\), far exceeding the corresponding first-shell intensities for PF$_6^{-}$ and TFSI$^{-}$, consistent with substantially stronger and more frequent direct \ce{Na+}--anion interactions in NaOTf relative to the more weakly coordinating anions.

Interestingly, UMA-OMol shows that solvent identity can significantly control these ion interactions. In particular, for the strongly coordinating OTf$^{-}$ anion, a substantially enhanced Na–OTf first shell peak and a steeper short-range growth in \(n(r)\) in DME than in DEGDME (Figure \ref{fgr:anion} c) demonstrates that the shorter glyme promotes stronger effective \ce{Na+}--OTf association, whereas the longer glyme more effectively competes for \ce{Na+} coordination and suppresses direct anion binding.\cite{geng2019influence,zhang2023research,chen2020ion,zhou2020engineering} 
The observed solvent–anion coupling effect, where solvent topology regulates ion-pairing even for strongly binding anions such as \ce{OTf-}, may enable strategic modulation of \ce{Na+} solvation structure and ion-association equilibria in Na-ion battery electrolytes.

\begin{figure*}[ht!]
\centering
  \includegraphics[height=5.0cm]{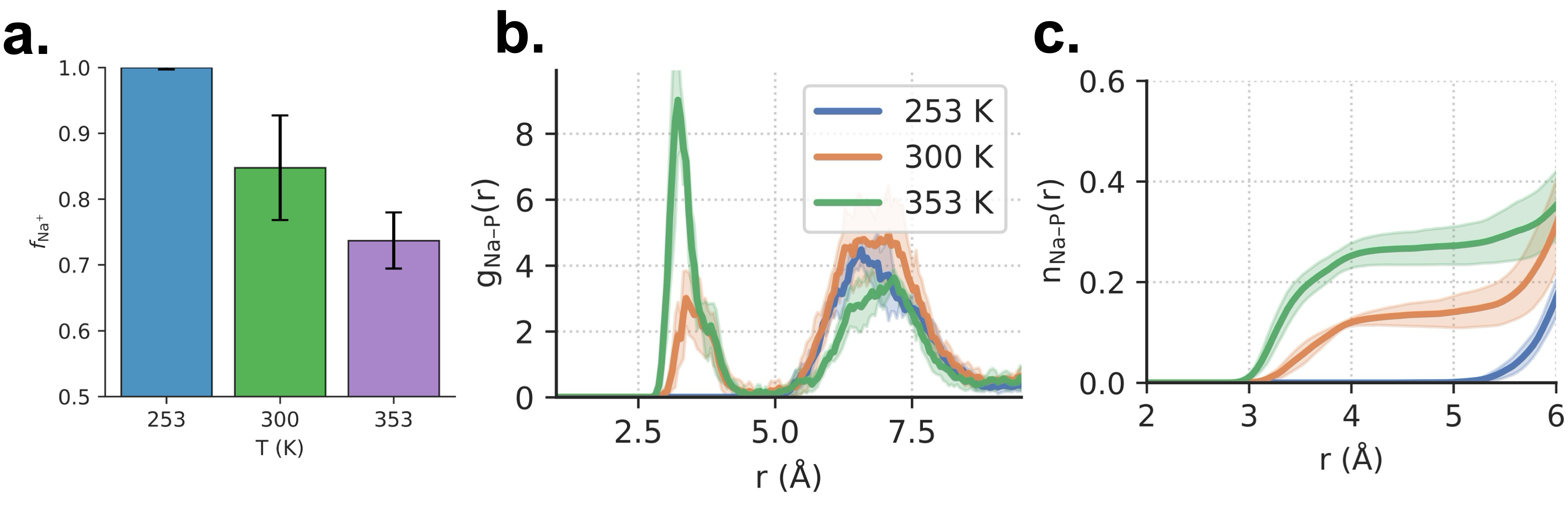}
\caption{Temperature dependence of NaPF$_6$ ion association in DME captured by UMA-OMol simulations. (a) Fraction of unpaired Na$^{+}$ ions at different temperatures. (b) Na–P (PF$_6^{-}$) radial distribution functions and (c) corresponding running coordination numbers at different temperatures. Error bars are obtained using the block-averaging methodology. The analysis reveals that all Na$^{+}$ ions are fully solvated at 253K, while at higher temperatures the cation solvation shell weakens and allows for increasing ion pair formation.}
\label{fgr:na_temp}
\end{figure*}

\paragraph{\ul{Temperature Effects}}

Temperature profoundly reorganizes the solvation landscape in both Li$^{+}$ and Na$^{+}$ liquid electrolytes, but with element-specific signatures reflecting their distinct ionic radii and binding energetics.\cite{han2024deciphering,wang2024temperature} To probe UMA-OMol's capacity to capture temperature-dependent changes in solvation structure, we studied 0.5 M \ce{NaPF6} and \ce{LiPF6} in DME at 253K, 300K, and 353K. At 253 K, both electrolytes exhibit highly ordered cation solvation, evidenced by sharp first-shell peaks in the $g_{\mathrm{Na-O(DME)}}(r)$ and $g_{\mathrm{Li-O(DME)}}(r)$ functions at $\sim$2.3 Å for Na$^{+}$ and $\sim$2.0 Å for Li$^{+}$ (Figures S5a, S6a), 100\% free ions observed in fractional analysis (Figures \ref{fgr:na_temp}a, S7a), and the complete absence of a first-shell peak in $g_{\mathrm{Na-P(PF_6)}}(r)$ and $g_{\mathrm{Li-P(PF_6)}}(r)$ functions (Figures \ref{fgr:na_temp}b, S7b), which trace cation--anion proximity. These data indicate tightly bound first-shell solvent coordination due to reduced thermal fluctuations at low-temperature conditions, which together render CIP formation a rare-event process on the presently sampled timescales. We anticipate that extending the simulation time or employing enhanced sampling strategies would allow occasional \ce{Na+}–\ce{PF6-} contact configurations to emerge, thereby more fully resolving the underlying free energy landscape governing ion association.

As temperature rises, the ion-solvent first-shell peaks attenuate and broaden, and average coordination decreases, signaling enhanced solvent exchange and weakened cation--solvent binding (Figures S5, S6). Concurrently, the Na--P radial distribution functions reveal an emerging first-shell contact peak near 3.0--3.3~\AA{}, with its intensity nearly doubling between 300~K and 353~K (Figure \ref{fgr:na_temp}b). The corresponding coordination integral $n_{\mathrm{Na-P}}(r)$ confirms a monotonic rise (Figure \ref{fgr:na_temp}c), and fraction analysis quantifies a significant decrease in the fraction of free ion (Figure \ref{fgr:na_temp}a) and an increase in ion-paired species---from negligible at 253~K to $\approx$ 0.27 at 353~K (Figure S8).\cite{lai2025linking,ringsby2021transport} Similar trends were exhibited for LiPF$_6$ (Figure S7), with one notable difference - the Li--P radial distribution function shows a distinct double first-shell peak, indicative of both monodentate and bidentate coordination motifs, as expected for Li$^{+}$.\cite{cresce2017solvation}

Overall, the thermal liberation of DME from the \ce{Na+} and \ce{Li+} solvation shells weakens the cation cage, enhances anion access, and shifts speciation toward contact and solvent shared ion pairs, yielding an entropy-driven reduction in the population of free \ce{Na+} or \ce{Li+} carriers despite decreasing viscosity.\cite{ringsby2021transport} These results show that UMA-OMol successfully captures the interplay between temperature-dependent enthalpic stabilization and entropic gains.

\paragraph{\ul{Concentration Effects}}
We next examined the correlations between electrolyte concentration and the propensity of ions to remain free or to form contact ion pairs, solvent separated pairs, and larger aggregates, each of which directly influences the effective transport of charge carriers especially under an applied potential.\cite{hu2022influence} Figure S9 summarizes the effect of NaPF$_6$ concentration in DME on the population of free Na$^+$ ions---defined here as cations without any PF$_6^-$ in their primary coordination shell---and on the fraction of Na$^+$ coordinated to exactly one PF$_6^-$ anion, between radial distances $3-10$ \AA{}. We observe an increased probability of Na$^+$--PF$_6^-$ CIPs followed by an expected decrease in the fraction of free \ce{Na+} ions with increasing electrolyte concentration. Further details are provided in Supplementary Information.

\begin{figure*}[ht!]
\centering
  \includegraphics[height=16cm]{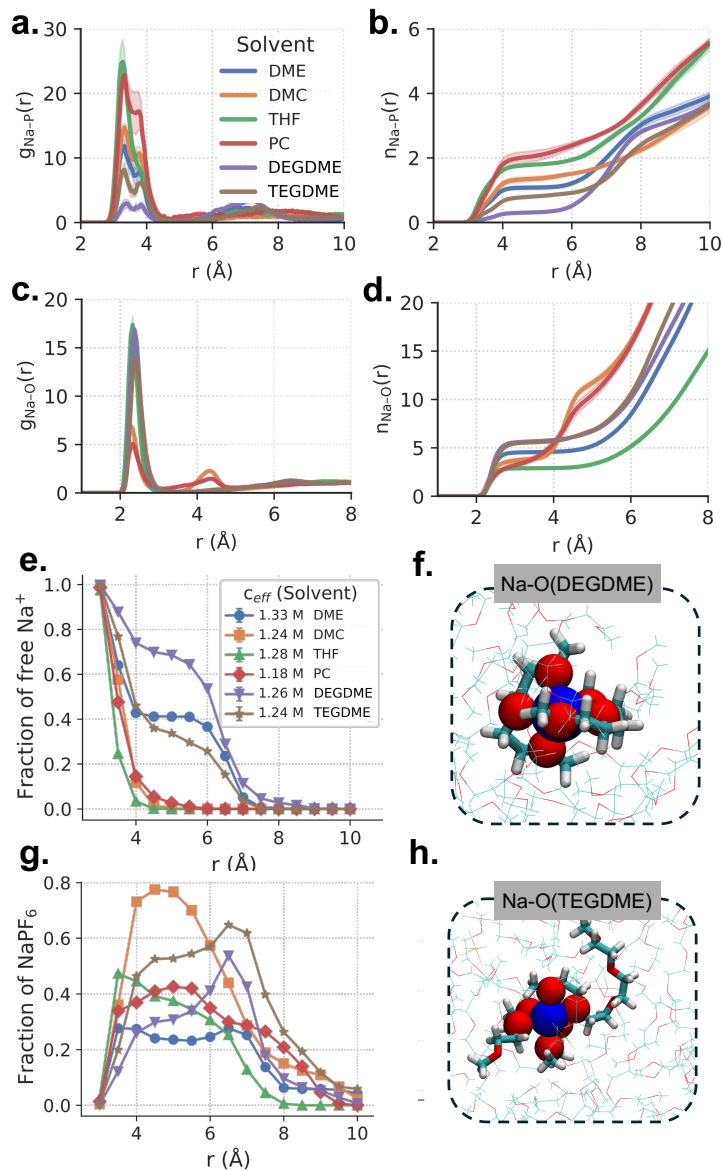}
\caption{Solvent effects on NaPF\(_6\) association in 1 M sodium electrolytes from UMA-OMol simulations:  
(a) Na–P radial distribution functions \(g_{\mathrm{Na-P}}(r)\) for dimethoxyethane (DME), dimethyl-carbonate (DMC), tetrahydrofuran (THF), propylene carbonate (PC), diglyme (DEGDME), and tetraethylene glycol dimethyl ether (TEGDME).  
(b) Corresponding Na–P running coordination numbers \(n_{\mathrm{Na-P}}(r)\).  
(c) Na–O radial distribution functions \(g_{\mathrm{Na-O}}(r)\) for the same solvent series.  
(d) Na–O running coordination numbers \(n_{\mathrm{Na-O}}(r)\).  
The variation in peak positions and coordination behavior highlights the strong solvent dependence of Na\(^{+}\)–PF\(_6^{-}\) association and Na\(^{+}\) solvation structure across linear glymes, cyclic ethers, and carbonate solvents. Comparison of DEGDME and TEGDME solvation of Na$^+$ in 1~M NaPF$_6$ electrolytes: (e) Fraction of free Na$^+$ ions versus radial cutoff distance. (g) Fraction of Na$^+$ forming CIPs. (f,h) Representative snapshots showing compact chelation by DEGDME excluding PF$_6^-$ versus open-chain TEGDME conformation enabling over-coordination and PF$_6^-$ insertion into the first solvation shell. }
  \label{fgr:solvent_effects}
\end{figure*}

\paragraph{\ul{Solvent Effects}}

Effective charge transport is strongly governed by the interactions between the charge carriers and their surrounding solvent environment.\cite{tian2022electrolyte,bergstrom2024ion,eshetu2020electrolytes} Therefore, we next investigated ion and solvent correlations across different solvent systems to quantify how variations in local solvation directly influence charge transport behavior.\cite{self2024solvation} Radial distribution functions and running coordination numbers provided in Figure \ref{fgr:solvent_effects}a-d, map how solvent identity modulates NaPF$_6$ association in different electrolyte solutions. In the Na--P radial distributions (Figure \ref{fgr:solvent_effects}a), the first-shell peak height decreases in the order PC~$\approx$~THF~$>$~DMC~$>$~DME~$>$~TEGDME~$>$~DEGDME. Integration of g(r) to 10~\AA{} (Figure \ref{fgr:solvent_effects}b) confirms that PC and THF stabilize the greatest number of contact/aggregate ion-pairs ($n_{\mathrm{Na-P}} \approx 5$--$6$), whereas DME, DEGDME and TEGDME limit Na--PF$_6$ CIPs to lower values. The complementary Na--O (DME) curves (Figure \ref{fgr:solvent_effects}c,d) show the inverse trend, where glymes furnish the densest oxygen solvation in the primary \ce{Na+} coordination shell. Thus, solvents with high donor number ether oxygens (DME, DEGDME, or TEGDME) most effectively compete with PF$_6^-$ for Na$^+$ coordination, yielding a larger population of “free” charge-carrying Na$^+$ ions (Figure \ref{fgr:solvent_effects}e), consistent with experimental trends.\cite{gutmann1976dn,morales2019ion,morales2021napf6carbonates,mandai2015solvatestability,ould2025napf6props}


Interestingly, the running coordination number profiles show that Na–O coordination is nearly identical in DEGDME and TEGDME (Figure \ref{fgr:solvent_effects}d), yet the extent of ion pairing differs substantially (Figure \ref{fgr:solvent_effects}b). Figures \ref{fgr:solvent_effects}e-g rationalize this behavior, demonstrating that while both solvents supply six ether oxygens to the primary solvation shell, differences in chain flexibility and local packing modulate the stabilization of contact and solvent shared ion pairs. Figure \ref{fgr:solvent_effects}e shows that the fraction of unpaired \ce{Na^{+}} in DEGDME decays more slowly, remaining above 65 \% (of total \ce{Na+}) at r = 5 Å, whereas TEGDME falls below 40 \% by the same radial distance from the \ce{Na+} centers. The complementary metric in Figure \ref{fgr:solvent_effects}g reveals a larger population, $> 20\%$ at r = 5 Å, of \ce{NaPF6} CIPs in TEGDME relative to DEGDME. Snapshots in Figure \ref{fgr:solvent_effects}f and \ref{fgr:solvent_effects}h clarify the microscopic origin, where DEGDME wraps into a compact, nearly octahedral chelate that sterically excludes \ce{PF_{6}^{-}}, favoring solvent separated or fully dissociated \ce{Na^{+}}. In contrast, the extended TEGDME chain adopts a flexible conformation that coordinates \ce{Na^{+}} while leaving void space that accommodates a \ce{PF_{6}^{-}} (Figure \ref{fgr:solvent_effects}f,h), thereby promoting CIP formation in the primary coordination shells. These conformational preferences demonstrate that subtle variations in solvent topology reshape the free-energy landscape of ion association, and that UMA-OMol captures the resulting redistribution among CIPs, SSIPs, and higher aggregates, with direct consequences for local solvation structure and charge transport.\cite{westman2018diglyme,jensen2020solvation,morales2019ion}


\paragraph{\ul{Solid/Liquid Interfacial Structure}}

\begin{figure}[ht!]
\centering
  \includegraphics[height=5.5cm]{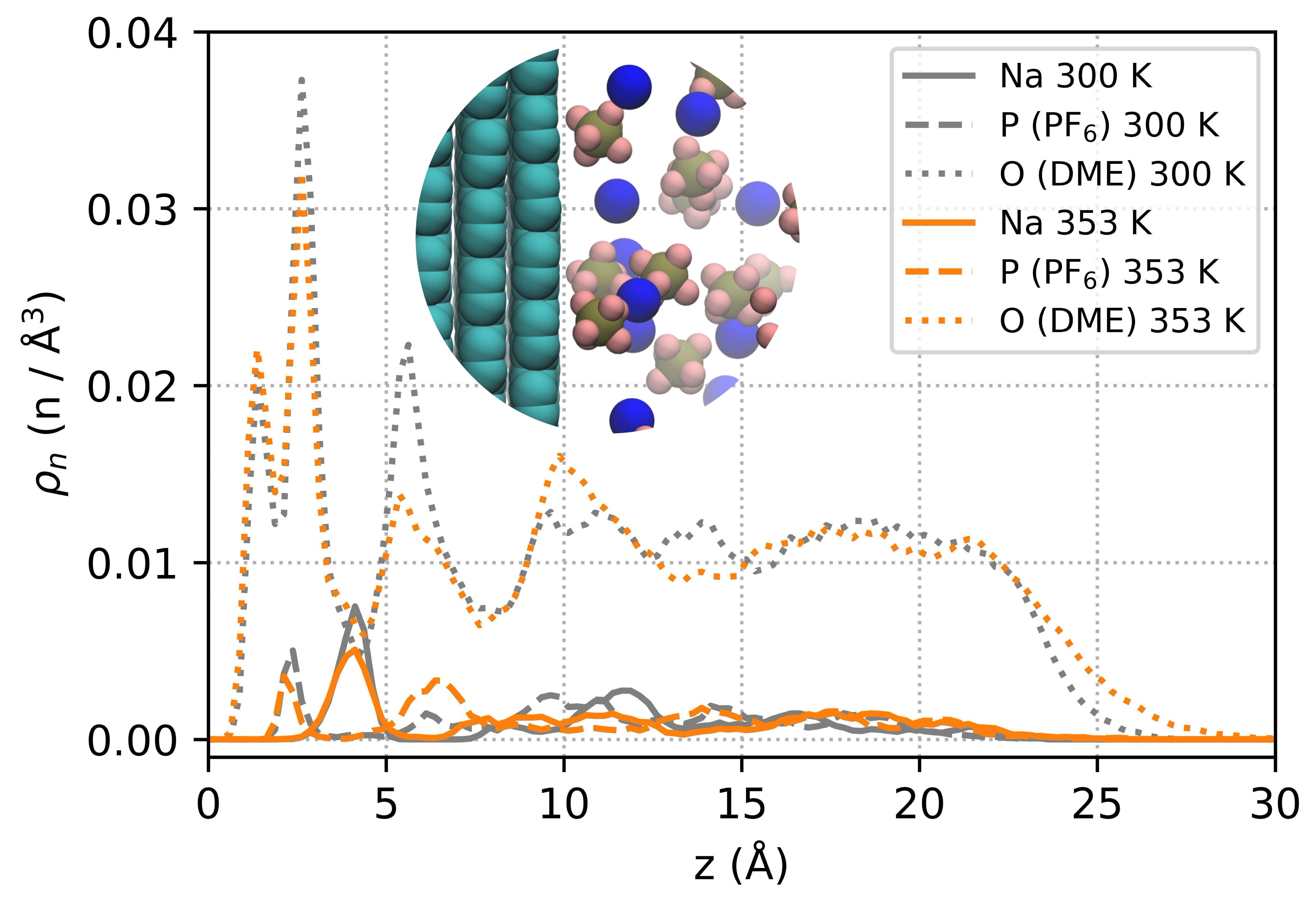}
\caption{Number density profiles ($\rho_{n}$) of \ce{Na+}, P (\ce{PF6-}), and O (DME) atoms along the surface normal (z) at the graphite–electrolyte interface for \ce{NaPF6} in DME at 300 K (gray) and 353 K (orange). At 300 K, pronounced solvent and ion layering is observed, with DME oxygens forming a compact interfacial layer that coordinates \ce{Na+} near the electrode surface, while \ce{PF6-} remains more weakly adsorbed. Elevated temperature reduces the amplitude of these oscillations, indicating thermal disruption of interfacial ordering and enhanced ion mobility. The inset shows a representative partial snapshot of the interfacial configuration highlighting solvent structuring adjacent to the graphite surface.}
  \label{fgr:interface}
\end{figure}

To assess the capability of UMA-OMol to capture electrolyte structuring at the electrode interface,\cite{finney2021electrochemistry,xu2019tailoring,kumar2024uranyl,zhang2022engineering} we simulated a graphite/DME system containing 1~M \ce{NaPF6} at 300~K and 353~K. A representative snapshot of the interfacial configuration is shown in Figure~S10. The number density profiles in Figure \ref{fgr:interface} reveal pronounced solvent and ion-layering at the graphite interface, indicating a highly ordered interfacial region. At 300~K, DME oxygen atoms form a compact adsorption layer that preferentially coordinates interfacial \ce{Na+} ions, while \ce{PF6-} anions exhibit weaker interfacial accumulation, consistent with solvent-mediated cation stabilization and limited anion contact with the electrode. The resulting oscillatory density profiles of oxygen, sodium, and phosphorus reflect a structured solvation environment in which \ce{Na+} resides within the first solvent shell coordinated primarily by interfacial oxygens, as shown in Figure~\ref{fgr:interface}. This organization stabilizes \ce{Na+} near the surface and constrains the initial electron transfer pathways that may initiate SEI formation.\cite{blau2021chemically} Upon increasing the temperature to 353~K, the amplitude of these density oscillations decreases and the profiles broaden, signaling enhanced thermal motion, partial desolvation, and weakened ion-solvent correlations within the interfacial region. Such interfacial disorder is known to promote molecular exchange between the interface and bulk, accelerating ion transport while simultaneously altering the accessibility and reductive stability of \ce{Na+} species at the electrode.\cite{zhang2024emerging,schott2024assess}

In this work, we benchmarked state-of-the-art large-scale MLIPs trained on inorganic materials DFT (Orb-OMat and SevenNet-OMat) and on the molecular DFT OMol25 dataset (UMA-OMol) against experimental observables to assess their reliability in describing the structure and thermodynamics of Na-ion battery electrolytes. Starting from the macroscopic density across a diverse set of organic liquids and electrolytes, parity plots of simulated versus experimental densities show that UMA-OMol achieves near-unity correlation ($R^2~>~0.98$) with minimal systematic bias, while MLIPs trained on inorganic materials consistently underpredict experimental values and exhibit larger scatter. X-ray structure factors $S(Q)$ for 1~M NaPF$_6$–diglyme demonstrate that UMA-OMol captures all principal experimental features, validating its accuracy in representing medium-range ordering. Small deviations observed at low $Q$ likely originate from long-range density correlations not captured by the MLIP given its 6 \AA{} cutoff. Difference plots, $\Delta S(Q)$, show that UMA-OMol exhibits the smallest residuals and minimal oscillatory behavior compared to SevenNet-OMat and Orb-OMat, confirming its superior fidelity in reproducing experimental scattering observables. Together, these results highlight how the chemically diverse and high-quality OMol25 molecular DFT training dataset increases MLIP prediction fidelity for intermolecular forces and improves thermodynamic consistency across liquid and electrolyte chemistries. 

Subsequent investigations using UMA-OMol to simulate Na-ion electrolyte systems with molecular dynamics while varying anion identity, temperature, concentration, solvent, and of a solid-liquid interface revealed atomistic insights useful to battery electrolyte design. Across the anion chemistries considered, OTf$^-$ displays substantially stronger association with Na$^+$ than \ce{PF6-} or \ce{TFSI-}, with ion pairing strengths strongly modulated by solvent topology through its control over primary coordination shell structure and competitive solvent binding. Our results demonstrate that UMA-OMol captures ion-pair hierarchies and binding strengths. Temperature-resolved simulations uncover the dynamic evolution of solvation and ion association. As temperature increases, Na$^+$–O(DME) coordination weakens while Na$^+$–P coordination strengthen, showing a transition from solvent-separated to contact or solvent-shared ion pairs. This structural reorganization parallels that observed in Li-based electrolytes but is more pronounced for Na due to its lower charge density and weaker solvation enthalpy. Finally, solvent topology emerges as a critical determinant of ionic association. Short-chain glymes (DME, diglyme) suppress NaPF$_6$ CIPs by stabilizing solvent-separated configurations, whereas conformationally flexible and extended TEGDME enhances over-coordination and ion aggregation. This solvent-dependent modulation of ion-pairing directly rationalizes the experimentally observed relationship between free-ion population and ionic conductivity. Similarly, our simulations revealed that UMA-OMol reproduces strong, temperature dependent structuring of electrolyte species at the graphite interface, with pronounced solvent and cation layering at room temperature that weakens at elevated temperature, driving migration of ions from the electrode-electrolyte interface to the bulk phase. 

OMol25 and UMA have demonstrably advanced MLIP capabilities for liquid electrolyte simulations; however, there remain important areas for future improvement. \emph{First}, while UMA-OMol substantially improves electrolyte MD stability compared to Orb-OMat and SevenNet-OMat, trajectories do still occasionally become unstable, even after equilibrating initial simulation boxes with a classical force field. Driving discovery with e.g. high-throughput screening will thus likely require either MLIP architectures with improved potential energy surface smoothness\cite{liu2026evaluation} or an automated MLIP-MD workflow infrastructure with robust on-the-fly error handling.\cite{blau2020accurate} \emph{Second}, electrolyte simulations in particular will benefit from MLIP architectures that go beyond short-range graph construction cutoffs -- e.g. via explicitly physics-informed approaches\cite{unke2019physnet,king2025machine} or global attention-based schemes\cite{qu2024importance,qu2026recipescalableattentionbasedmlips} -- to accurately describe long-range electrostatics while still effectively scaling to tens of thousands of atoms. \emph{Third}, simulating transport properties in liquid electrolytes, and even more so in polymer electrolytes,\cite{levine2025openpol} necessitates reaching MD timescales that are out of reach for current large-scale pre-trained MLIPs. While improving GPU hardware will help, model distillation also shows great promise,\cite{amin2025towards} provided that a standardized and automated procedure can be established that marries pre-trained model accuracy with at least an order of magnitude inference speedup. \emph{Fourth}, it remains unclear if range-separated hybrid DFT, as was used to construct OMol25, is sufficient to train MLIPs that predict experimental observables within chemical accuracy. It may be that electronic structure data beyond DFT accuracy will be required for model fine-tuning, though generating such data spanning the same chemical diversity as OMol25 will require massive computational resources. \emph{Finally}, we see great potential in more closely marrying high-quality experimental data with MLIP-based atomistic simulations. Establishing comprehensive experimental benchmarks with broad chemical diversity and varied observables that can be directly compared to atomistic simulations will quantify model accuracy as architectures and training data improve and facilitate hybrid approaches like digital twins\cite{qian2025digital} and multi-modal learning on both experimental and ab-initio data.\cite{gumber2025going,raja2024stability}

\section*{Supplementary Information}
Methodology, system densities, structure factors, radial distribution functions and coordination number, fraction of free ions, and raw data for all experimental measurements.

\section*{Acknowledgments}

All authors acknowledge support from the Energy Storage Research Alliance (ESRA) (DE-AC02-06CH11357), an Energy Innovation Hub funded by the U.S. Department of Energy, Office of Science, Basic Energy Sciences. This work used computational resources provided by the National Energy Research Scientific Computing Center (NERSC), a U.S. Department of Energy Office of Science User Facility operated under Contract DE-AC02-05CH11231, and the Lawrencium computational cluster resource provided by the IT Division at Lawrence Berkeley National Laboratory (Supported by the Director, Office of Science, Office of Basic Energy Sciences, of the U.S. Department of Energy under Contract No. DE-AC02-05CH11231). This research was performed on APS beam time award(s) (DOI: 10.46936/APS-192520/60015964) from the Advanced Photon Source, a U.S. Department of Energy (DOE) Office of Science user facility operated for the DOE Office of Science by Argonne National Laboratory under Contract No. DE-AC02-06CH11357.

\section*{Conflicts of Interest}
None.


\bibliographystyle{achemso}
\bibliography{achemso-demo}

\end{document}


\maketitle

\section*{Methodology}

\subsection{Simulation Protocol}

Initial simulation coordinates were generated using \textsc{Packmol}.\cite{martinez2009packmol} A relatively large $30$~{\AA} simulation cell edge length was chosen to prevent any finite-size effects. Solvent molecules were added to match their experimental densities. Each system was first annealed and equilibrated in the isothermal--isobaric (NPT) ensemble using classical OPLS-AA force fields.\cite{kaminski2001evaluation,jorgensen2005potential} These equilibrated configurations served as the starting point for the machine learning interatomic potential molecular dynamics (MLIP-MD) simulations. MLIP-MD simulations were performed using the \textsc{Atomic Simulation Environment} (ASE) framework in combination with different MLIP calculators.\cite{larsen2017atomic} Atomic interactions were modeled using the \texttt{uma-s-1p1} neural network potential, OMol task, via the \texttt{FAIRChemCalculator}.\cite{wood2025family} For comparison, the \texttt{orb-v3-conservative-inf-omat} model was used with the \texttt{OrbCalculator}, and the \texttt{SevenNet-omat} model was employed with the \texttt{SevenNetCalculator}. MLIP-MD simulations were carried out in the NPT ensemble at 300~K and 1~bar, with periodic boundary conditions applied in all three spatial dimensions. Temperature was changed from 253 K to 353 K to study the temperature effects. A time step of 1~fs was used to integrate the equations of motion. Thermostat and barostat relaxation times were set to 100~fs and a pressure coupling factor of 0.1, respectively. Trajectories were saved every 10~fs, and each simulation was run for a minimum of 0.5~ns per system, where the final 0.2 ns were used for data analysis. 

The fraction of free and ion-paired \ce{Na+} species was quantified from MD trajectories using a coordination-based counting analysis of \ce{Na-P} interactions. For each saved frame of the trajectory, all sodium (\ce{Na}) and phosphorus (\ce{P}) atoms were first identified, and pairwise \ce{Na-P} distances were computed under periodic boundary conditions using the minimum-image convention to ensure accurate treatment of ions crossing the simulation box boundaries. A continuous set of radial cutoffs \(r\) ranging from 0.05 to 8.0~\AA\ in increments of 0.05~\AA\ was applied to determine the number of \ce{PF6-} anions surrounding each \ce{Na+} within a given distance. For each cutoff \(r\), the number of sodium ions having \textit{exactly one} phosphorus atom within \(r\) was counted, yielding the population of \ce{Na+} that forms a distinct one-to-one \ce{Na-PF6} contact ion pair (CIP). Dividing this quantity by the total number of \ce{Na+} ions in the system gives the cumulative fraction \(f(r)\) of ion-paired \ce{Na+} as a function of distance. Similarly, the number of \ce{Na+} with no phosphorus atom within \(r\) corresponded to the population of "free" \ce{Na+} ions that are not directly paired with a \ce{PF6-} anion within the specified cutoff. Block averaging methodology was used to obtain error bars.

The X-ray total scattering structure factor \( S(Q) \) of the \ce{NaPF6}–diglyme electrolyte was computed from the MD trajectories using the \textsc{TRAVIS} program.\cite{brehm2020travis} The simulation trajectory was first converted into the TRAVIS-readable format, ensuring the correct atomic numbers, masses, and scattering form factors were assigned to each element. The program evaluates the partial pair distribution functions \( g_{\alpha\beta}(r) \) between atom types \(\alpha\) and \(\beta\), which describe the normalized probability of finding a pair of atoms separated by distance \(r\). The total X-ray structure factor is then obtained by Fourier transformation of these pair distribution functions according to the Debye scattering formalism implemented in TRAVIS using formula:

\[
S(Q) = 1 + \frac{4\pi}{\langle b \rangle^2 V} 
\sum_{\alpha} \sum_{\beta} c_{\alpha} c_{\beta} b_{\alpha} b_{\beta}
\int_0^{r_{\text{max}}} \left[g_{\alpha\beta}(r) - 1\right]
\frac{\sin(Qr)}{Qr} \, r^2 \, dr,
\]

where \( Q \) is the scattering vector magnitude, \( V \) is the simulation cell volume, \( c_{\alpha} \) and \( c_{\beta} \) are the atomic concentrations, and \( b_{\alpha} \) and \( b_{\beta} \) are the X-ray scattering factors for the atomic species \(\alpha\) and \(\beta\). The angle brackets denote the weighted average over all atom types, where, 
\(\langle b \rangle = \sum_{\gamma} c_{\gamma} b_{\gamma}\).

\subsection{Experimental Details}

\textbf{Densimetry}

Density measurements were performed using an Anton Paar DMA 4500 M oscillating u-tube instrument, calibrated to dry air and ISO 17034-certified water, with bubble-free filling of the tube confirmed by the onboard camera system. Measurements were taken at temperatures from 273.2 K to 353.2 K in 5 K increments, regulated with the onboard Peltier elements, and dry \ce{N2} gas was plumbed into the chamber to prevent condensation on the exterior of the tube at low temperature. The u-tube was flushed 3x rapidly with 100\% ethanol between samples and the tube was fully dried with the \ce{N2} gas pump (600 s of flushing). GxP air and water density checks were taken between variable-temperature data collection runs to confirm the calibration (followed again by ethanol flushing and complete drying before injecting the next sample).

\noindent \textbf{X-ray Total Scattering Measurements}

X-ray total scattering measurements were performed at the Advanced Photon Source on beamline 11-ID-B. High-energy X-rays with a wavelength of $\lambda = 0.2099$ \AA\ were used for all measurements. Electrolyte samples were loaded into 5 mm NMR tubes inside an argon-filled glovebox to prevent exposure to air and moisture. The tubes were sealed inside the glovebox using Apiezon Q sealing grease and autosampler caps to maintain an inert environment during measurement. Two-dimensional scattering patterns were collected using a PerkinElmer (PE) area detector.\cite{chupas2003rapid,chupas2007applications} The raw detector images were azimuthally integrated using GSAS-II to obtain one-dimensional scattering intensity profiles $I(Q)$ as a function of momentum transfer.\cite{hammersley1996two,toby2013gsas} An empty tube background and instrument background were measured under identical experimental conditions and subtracted during the data reduction procedure. The integrated data were subsequently processed using xPDFsuite.\cite{juhas2013pdfgetx3} Standard corrections were applied, including background subtraction, container scattering removal, Compton scattering correction, and normalization to obtain the total X-ray structure factor $S(Q)$. The resulting structure factors were used for comparison with simulated scattering patterns generated from molecular dynamics trajectories. The experimental configuration provided a usable momentum transfer range up to approximately $Q \approx 21.5
$ \AA$^{-1}$, enabling characterization of short- and intermediate-range structural correlations in the electrolyte solutions.

\section{Results and Discussion}

\begin{table}[htbp]
\centering
\caption{Comparison of simulated densities (g/cm$^3$) from UMA-OMol, SevenNet-OMat, and Orb-OMat models against experimental values at 298.2~K.}
\label{tab:density_comparison}
\begin{tabular}{lcccc}
\hline\hline
System & UMA-OMol & SevenNet-OMat & Orb-OMat & Expt. \\
\hline
\multicolumn{5}{l}{\textit{Pure Solvents}} \\
Pure DME & 0.957 & 0.777 & 0.781 & 0.861 \\
Pure DEGDME & 1.042 & 0.840 & 0.893 & 0.939 \\
Pure TEGDME & 1.108 & --- & --- & 1.006 \\
Pure PC & 1.258 & 0.949 & 1.182 & 1.199 \\
Pure DEG & 1.213 & 1.021 & 0.893 & 1.113 \\
Pure DMC & 1.157 & 0.943 & 1.383 & 1.063 \\
\hline
\multicolumn{5}{l}{\textit{0.1~M Electrolytes}} \\
0.1 M NaOTf in DME & 0.966 & --- & --- & 0.873 \\
0.1 M NaPF6 in DME & 0.969 & --- & --- & 0.873 \\
0.1 M NaOTf in DEGDME & 1.056 & --- & --- & 0.950 \\
0.1 M NaPF6 in DEGDME & 1.045 & --- & --- & 0.952 \\
0.1 M NaOTf in TEGDME & 1.119 & --- & --- & 1.016 \\
0.1 M NaPF6 in TEGDME & 1.116 & --- & --- & 1.014 \\
0.1 M NaOTf in PC & 1.272 & --- & --- & 1.207 \\
0.1 M NaPF6 in PC & 1.267 & --- & --- & 1.206 \\
\hline
\multicolumn{5}{l}{\textit{0.5~M Electrolytes}} \\
0.5 M LiPF6 in DME & 1.031 & --- & --- & 0.909 \\
\hline
\multicolumn{5}{l}{\textit{1.0~M Electrolytes}} \\
1.0 M KPF6 in DME & 1.064 & --- & --- & 1.001 \\
1.0 M NaOTf in DME & 1.077 & 0.896 & 0.620 & 0.980 \\
1.0 M NaPF6 in DME & 1.106 & 0.910 & 0.940 & 0.995 \\
1.0 M NaTFSI in DME & 1.146 & 0.983 & 0.938 & 1.055 \\
1.0 M NaOTf in DEGDME & 1.148 & --- & --- & 1.047 \\
1.0 M NaPF6 in DEGDME & 1.166 & 0.670 & 0.982 & 1.059 \\
1.0 M NaOTf in TEGDME & 1.219 & --- & --- & 1.106 \\
1.0 M NaPF6 in TEGDME & 1.214 & --- & --- & 1.115 \\
1.0 M NaOTf in PC & 1.372 & --- & --- & 1.277 \\
1.0 M NaPF6 in PC & 1.377 & 1.024 & 1.261 & 1.291 \\
\hline\hline
\end{tabular}
\end{table}

\begin{table}[htbp]
\centering
\caption{Density (g/cm$^3$) of 0.5\,M LiPF$_6$ and NaPF$_6$ in DME: simulation (UMA) vs.\ experiment.}
\label{tab:density_sim_exp}
\begin{tabular}{c cc cc}
\hline\hline
Temperature & \multicolumn{2}{c}{0.5\,M LiPF$_6$/DME} & \multicolumn{2}{c}{0.5\,M NaPF$_6$/DME} \\
\cline{2-3} \cline{4-5}
(K) & Sim. & Exp. & Sim. & Exp. \\
\hline
  253 & 1.075 & --- & 1.080 & --- \\
  273.15 & 1.067 & 0.952 & 1.059 & 0.962 \\
  278.15 & --- & 0.946 & --- & 0.957 \\
  283.15 & --- & 0.941 & --- & 0.951 \\
  288.15 & --- & 0.936 & --- & 0.946 \\
  293.15 & --- & 0.930 & --- & 0.940 \\
  298.15 & --- & 0.924 & --- & 0.935 \\
  300 & 1.031 & --- & 1.043 & --- \\
  303.15 & --- & 0.919 & --- & 0.929 \\
  308.15 & --- & 0.913 & --- & 0.923 \\
  313.15 & --- & 0.907 & --- & 0.918 \\
  318.15 & --- & 0.901 & --- & 0.912 \\
  323.15 & 1.010 & 0.895 & 1.011 & 0.906 \\
  328.15 & --- & 0.889 & --- & 0.900 \\
  333.15 & --- & 0.883 & --- & 0.895 \\
  338.15 & --- & 0.877 & --- & 0.889 \\
  343.15 & --- & 0.871 & --- & 0.883 \\
  348.15 & --- & 0.865 & --- & 0.877 \\
  353.15 & 0.971 & 0.859 & 0.993 & 0.871 \\
\hline\hline
\end{tabular}
\end{table}

\begin{longtable}{ccccccc}
\caption{Experimental structure factors $S(Q)$ for 1\,M Na-ion electrolyte systems.}
\label{tab:sq} \\
\toprule
$Q$ (\AA$^{-1}$) & \rotatebox{45}{NaPF$_6$/DME} & \rotatebox{45}{NaPF$_6$/DEGDME} & \rotatebox{45}{NaPF$_6$/TEGDME} & \rotatebox{45}{NaOTf/DME} & \rotatebox{45}{NaOTf/DEGDME} & \rotatebox{45}{NaTFSI/DME} \\
\midrule
\endfirsthead
\toprule
$Q$ (\AA$^{-1}$) & \rotatebox{45}{NaPF$_6$/DME} & \rotatebox{45}{NaPF$_6$/DEGDME} & \rotatebox{45}{NaPF$_6$/TEGDME} & \rotatebox{45}{NaOTf/DME} & \rotatebox{45}{NaOTf/DEGDME} & \rotatebox{45}{NaTFSI/DME} \\
\midrule
\endhead
\midrule
\multicolumn{7}{r}{\textit{Continued on next page}} \\
\endfoot
\bottomrule
\endlastfoot
1.2070 & 1.2698 & 1.1661 & 0.9872 & 1.1006 & 0.9404 & 1.0697 \\
1.4028 & 2.2848 & 2.5381 & 2.4992 & 2.1618 & 2.4536 & 1.7924 \\
1.5986 & 2.8286 & 2.9435 & 2.9030 & 2.4155 & 2.6462 & 2.0700 \\
1.7945 & 1.3930 & 1.3499 & 1.3029 & 1.5942 & 1.5139 & 1.3370 \\
1.9903 & 0.6862 & 0.6596 & 0.6763 & 1.0588 & 0.9466 & 0.9394 \\
2.1861 & 0.4710 & 0.4724 & 0.5076 & 0.8230 & 0.7300 & 0.8146 \\
2.3819 & 0.4440 & 0.4738 & 0.5022 & 0.7161 & 0.6704 & 0.8112 \\
2.5777 & 0.4998 & 0.5590 & 0.5608 & 0.6826 & 0.6920 & 0.8601 \\
2.7735 & 0.6174 & 0.6997 & 0.7039 & 0.7548 & 0.7977 & 0.9570 \\
2.9693 & 0.7699 & 0.8572 & 0.8924 & 0.8932 & 0.9417 & 1.0292 \\
3.1651 & 0.9102 & 0.9689 & 0.9880 & 0.9466 & 0.9855 & 1.0140 \\
3.3609 & 0.9971 & 1.0037 & 0.9755 & 0.9025 & 0.9191 & 0.9397 \\
3.5568 & 1.0373 & 0.9920 & 0.9261 & 0.8223 & 0.8123 & 0.8373 \\
3.7526 & 1.0349 & 0.9607 & 0.8918 & 0.7528 & 0.7222 & 0.7331 \\
3.9484 & 1.0005 & 0.9238 & 0.8861 & 0.7302 & 0.6878 & 0.6692 \\
4.1442 & 0.9632 & 0.8968 & 0.8987 & 0.7637 & 0.7159 & 0.6611 \\
4.3400 & 0.9363 & 0.8860 & 0.9230 & 0.8381 & 0.7917 & 0.7148 \\
4.5358 & 0.9310 & 0.9006 & 0.9575 & 0.9240 & 0.8930 & 0.8224 \\
4.7316 & 0.9423 & 0.9335 & 0.9914 & 0.9938 & 0.9891 & 0.9426 \\
4.9274 & 0.9738 & 0.9869 & 1.0384 & 1.0745 & 1.0932 & 1.0529 \\
5.1232 & 1.0398 & 1.0697 & 1.1047 & 1.1780 & 1.2042 & 1.1550 \\
5.3191 & 1.1609 & 1.1876 & 1.1991 & 1.3035 & 1.3219 & 1.2981 \\
5.5149 & 1.3019 & 1.3084 & 1.3047 & 1.4085 & 1.4178 & 1.4671 \\
5.7107 & 1.3987 & 1.3805 & 1.3662 & 1.4297 & 1.4425 & 1.5218 \\
5.9065 & 1.4097 & 1.3742 & 1.3509 & 1.3624 & 1.3798 & 1.4267 \\
6.1023 & 1.3174 & 1.2804 & 1.2458 & 1.2317 & 1.2470 & 1.2465 \\
6.2981 & 1.1407 & 1.1182 & 1.0628 & 1.0593 & 1.0676 & 1.0496 \\
6.4939 & 0.9299 & 0.9299 & 0.8593 & 0.8932 & 0.8892 & 0.8865 \\
6.6897 & 0.7335 & 0.7581 & 0.6844 & 0.7559 & 0.7415 & 0.7780 \\
6.8855 & 0.5904 & 0.6334 & 0.5677 & 0.6636 & 0.6449 & 0.7199 \\
7.0814 & 0.5095 & 0.5618 & 0.5116 & 0.6167 & 0.5948 & 0.6980 \\
7.2772 & 0.4928 & 0.5443 & 0.5131 & 0.6080 & 0.5839 & 0.6935 \\
7.4730 & 0.5336 & 0.5769 & 0.5812 & 0.6454 & 0.6222 & 0.7112 \\
7.6688 & 0.6241 & 0.6536 & 0.6890 & 0.7091 & 0.6908 & 0.7317 \\
7.8646 & 0.7544 & 0.7671 & 0.8144 & 0.7770 & 0.7680 & 0.7460 \\
8.0604 & 0.9044 & 0.9017 & 0.9519 & 0.8553 & 0.8581 & 0.7810 \\
8.2562 & 1.0481 & 1.0328 & 1.0793 & 0.9416 & 0.9532 & 0.8413 \\
8.4520 & 1.1708 & 1.1453 & 1.1753 & 1.0158 & 1.0332 & 0.9048 \\
8.6478 & 1.2490 & 1.2162 & 1.2528 & 1.0983 & 1.1150 & 0.9883 \\
8.8437 & 1.2878 & 1.2501 & 1.2781 & 1.1475 & 1.1576 & 1.0508 \\
9.0395 & 1.2799 & 1.2407 & 1.2701 & 1.1815 & 1.1859 & 1.1122 \\
9.2353 & 1.2428 & 1.2069 & 1.2109 & 1.1676 & 1.1651 & 1.1347 \\
9.4311 & 1.1808 & 1.1508 & 1.1401 & 1.1482 & 1.1412 & 1.1595 \\
9.6269 & 1.1089 & 1.0887 & 1.0591 & 1.1161 & 1.1033 & 1.1661 \\
9.8227 & 1.0387 & 1.0286 & 0.9923 & 1.0929 & 1.0793 & 1.1736 \\
10.0185 & 0.9855 & 0.9861 & 0.9544 & 1.0874 & 1.0734 & 1.1832 \\
10.2143 & 0.9597 & 0.9698 & 0.9431 & 1.0933 & 1.0819 & 1.1860 \\
10.4101 & 0.9472 & 0.9681 & 0.9518 & 1.0957 & 1.0867 & 1.1626 \\
10.6060 & 0.9468 & 0.9753 & 0.9730 & 1.0944 & 1.0922 & 1.1319 \\
10.8018 & 0.9467 & 0.9751 & 0.9817 & 1.0675 & 1.0673 & 1.0754 \\
10.9976 & 0.9368 & 0.9595 & 0.9590 & 1.0048 & 1.0062 & 0.9992 \\
11.1934 & 0.9232 & 0.9320 & 0.9226 & 0.9349 & 0.9315 & 0.9291 \\
11.3892 & 0.9075 & 0.9028 & 0.8838 & 0.8645 & 0.8583 & 0.8677 \\
11.5850 & 0.8852 & 0.8712 & 0.8555 & 0.8099 & 0.8013 & 0.8209 \\
11.7808 & 0.8702 & 0.8541 & 0.8429 & 0.7846 & 0.7756 & 0.7933 \\
11.9766 & 0.8564 & 0.8410 & 0.8356 & 0.7715 & 0.7625 & 0.7705 \\
12.1724 & 0.8515 & 0.8395 & 0.8394 & 0.7773 & 0.7692 & 0.7599 \\
12.3683 & 0.8623 & 0.8539 & 0.8502 & 0.7976 & 0.7912 & 0.7618 \\
12.5641 & 0.8704 & 0.8676 & 0.8647 & 0.8239 & 0.8219 & 0.7777 \\
12.7599 & 0.8895 & 0.8909 & 0.8880 & 0.8574 & 0.8568 & 0.8092 \\
12.9557 & 0.9142 & 0.9202 & 0.9236 & 0.8987 & 0.8990 & 0.8574 \\
13.1515 & 0.9696 & 0.9762 & 0.9740 & 0.9592 & 0.9622 & 0.9281 \\
13.3473 & 1.0135 & 1.0217 & 1.0304 & 1.0206 & 1.0297 & 1.0024 \\
13.5431 & 1.0750 & 1.0819 & 1.0777 & 1.0705 & 1.0803 & 1.0651 \\
13.7389 & 1.1125 & 1.1162 & 1.1109 & 1.1130 & 1.1213 & 1.1141 \\
13.9347 & 1.1432 & 1.1425 & 1.1298 & 1.1411 & 1.1515 & 1.1513 \\
14.1306 & 1.1337 & 1.1306 & 1.1189 & 1.1451 & 1.1445 & 1.1613 \\
14.3264 & 1.1174 & 1.1095 & 1.0931 & 1.1339 & 1.1309 & 1.1565 \\
14.5222 & 1.0779 & 1.0701 & 1.0572 & 1.1134 & 1.1023 & 1.1413 \\
14.7180 & 1.0263 & 1.0196 & 1.0004 & 1.0668 & 1.0516 & 1.0951 \\
14.9138 & 0.9790 & 0.9740 & 0.9622 & 1.0355 & 1.0192 & 1.0655 \\
15.1096 & 0.9409 & 0.9382 & 0.9315 & 0.9995 & 0.9838 & 1.0244 \\
15.3054 & 0.9191 & 0.9183 & 0.9105 & 0.9686 & 0.9557 & 0.9904 \\
15.5012 & 0.9165 & 0.9179 & 0.9144 & 0.9564 & 0.9404 & 0.9685 \\
15.6970 & 0.9177 & 0.9203 & 0.9246 & 0.9356 & 0.9293 & 0.9399 \\
15.8929 & 0.9232 & 0.9267 & 0.9454 & 0.9303 & 0.9270 & 0.9268 \\
16.0887 & 0.9336 & 0.9370 & 0.9380 & 0.8972 & 0.9039 & 0.8877 \\
16.2845 & 0.9460 & 0.9453 & 0.9474 & 0.8953 & 0.9017 & 0.8758 \\
16.4803 & 0.9486 & 0.9446 & 0.9530 & 0.8962 & 0.8988 & 0.8685 \\
16.6761 & 0.9554 & 0.9483 & 0.9530 & 0.9029 & 0.9019 & 0.8734 \\
16.8719 & 0.9575 & 0.9497 & 0.9395 & 0.8953 & 0.8967 & 0.8666 \\
17.0677 & 0.9602 & 0.9516 & 0.9552 & 0.9181 & 0.9165 & 0.8976 \\
17.2635 & 0.9736 & 0.9637 & 0.9497 & 0.9280 & 0.9254 & 0.9172 \\
17.4593 & 0.9799 & 0.9729 & 0.9550 & 0.9443 & 0.9467 & 0.9461 \\
17.6552 & 0.9768 & 0.9717 & 0.9687 & 0.9743 & 0.9695 & 0.9778 \\
17.8510 & 0.9762 & 0.9739 & 0.9703 & 0.9881 & 0.9848 & 1.0112 \\
18.0468 & 0.9855 & 0.9854 & 0.9743 & 1.0095 & 0.9999 & 1.0291 \\
18.2426 & 0.9755 & 0.9791 & 0.9812 & 1.0233 & 1.0201 & 1.0519 \\
18.4384 & 0.9740 & 0.9834 & 0.9880 & 1.0355 & 1.0350 & 1.0571 \\
18.6342 & 0.9877 & 0.9968 & 0.9919 & 1.0347 & 1.0385 & 1.0581 \\
18.8300 & 0.9878 & 0.9980 & 0.9929 & 1.0410 & 1.0390 & 1.0478 \\
19.0258 & 0.9941 & 1.0035 & 0.9990 & 1.0381 & 1.0396 & 1.0442 \\
19.2217 & 1.0081 & 1.0142 & 0.9910 & 1.0107 & 1.0096 & 1.0060 \\
19.4175 & 1.0059 & 1.0034 & 0.9945 & 1.0020 & 1.0028 & 0.9904 \\
19.6133 & 1.0083 & 1.0014 & 0.9955 & 0.9992 & 0.9855 & 0.9873 \\
19.8091 & 1.0040 & 0.9916 & 0.9781 & 0.9685 & 0.9593 & 0.9616 \\
\end{longtable}

\subsection{Concentration Effects}

Figure S9, Panel~a shows that the free-ion fraction (within r $<$ 5 Å) decreases sharply with increasing electrolyte concentration (from 0.66 M to 3.14 M). At the lowest effective concentration (0.66~M), more than 65\% of Na$^+$ remain free out to $r \approx 5$~\AA{}, and this fraction only drops to zero near 8~\AA{}. At higher concentrations, competitive coordination by PF$_6^-$ rapidly displaces solvent molecules, reducing the free-ion population to $\sim$10\% by $r = 4$~\AA{} at 3.14~M. This trend reflects the increased probability of Na$^+$--PF$_6^-$ contact formation as ionic strength rises, which in turn may lower the number of highly mobile, conductivity-relevant Na$^+$ carriers. 

Figure S9, Panel~b quantifies the fraction of Na$^+$ forming exactly one Na--PF$_6$ contact. At short distances ($r \lesssim 4$~\AA{}), these correspond to CIPs, and their population grows markedly with concentration---rising from $\sim$0.18 at 0.66~M to $\sim$0.5 at 3.14~M. Beyond $r \approx 5.5$--7~\AA{}, the curves capture SSIPs. At low concentration, the SSIP fraction is significant and peaks near 7~\AA{}, indicating stable, well-defined single-anion associations. As concentration increases, this SSIP peak diminishes and broadens, consistent with the formation of higher-order ion-aggregates in which Na$^+$ is coordinated to multiple PF$_6^-$ anions at longer-range rather than maintaining a one-to-one pairing, showing ability of OMol trained MLIP to capture concentration effects on ion solvation structure across a diverse concentration range.\cite{geng2019influence, dhir2023fundamental, zarayeneh2022dynamic, monti2020towards, wahlers2016solvation}

\newpage

\begin{figure}[ht!]
\centering
    \captionsetup{name= Figure }
  \includegraphics[height=7cm]{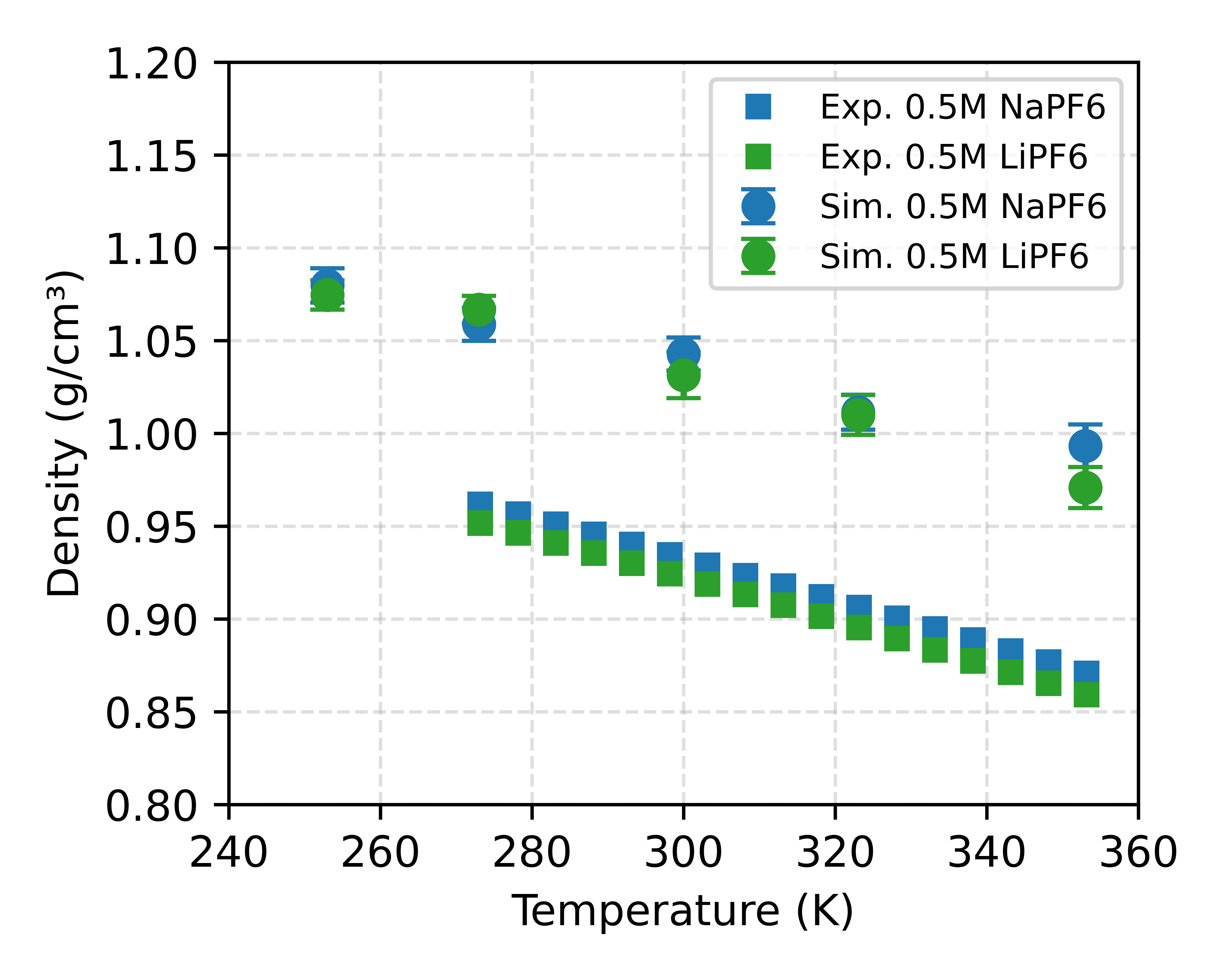}
\caption{Experimental and UMA-OMol simulated densities of \ce{LiPF6} or \ce{NaPF6} electrolytes in DME (0.5 M initial concentration) at different temperatures.  }

  \label{fgr:area}
\end{figure}

\begin{figure*}[ht!]
\centering
    \captionsetup{name= Figure }
  \includegraphics[height=10cm]{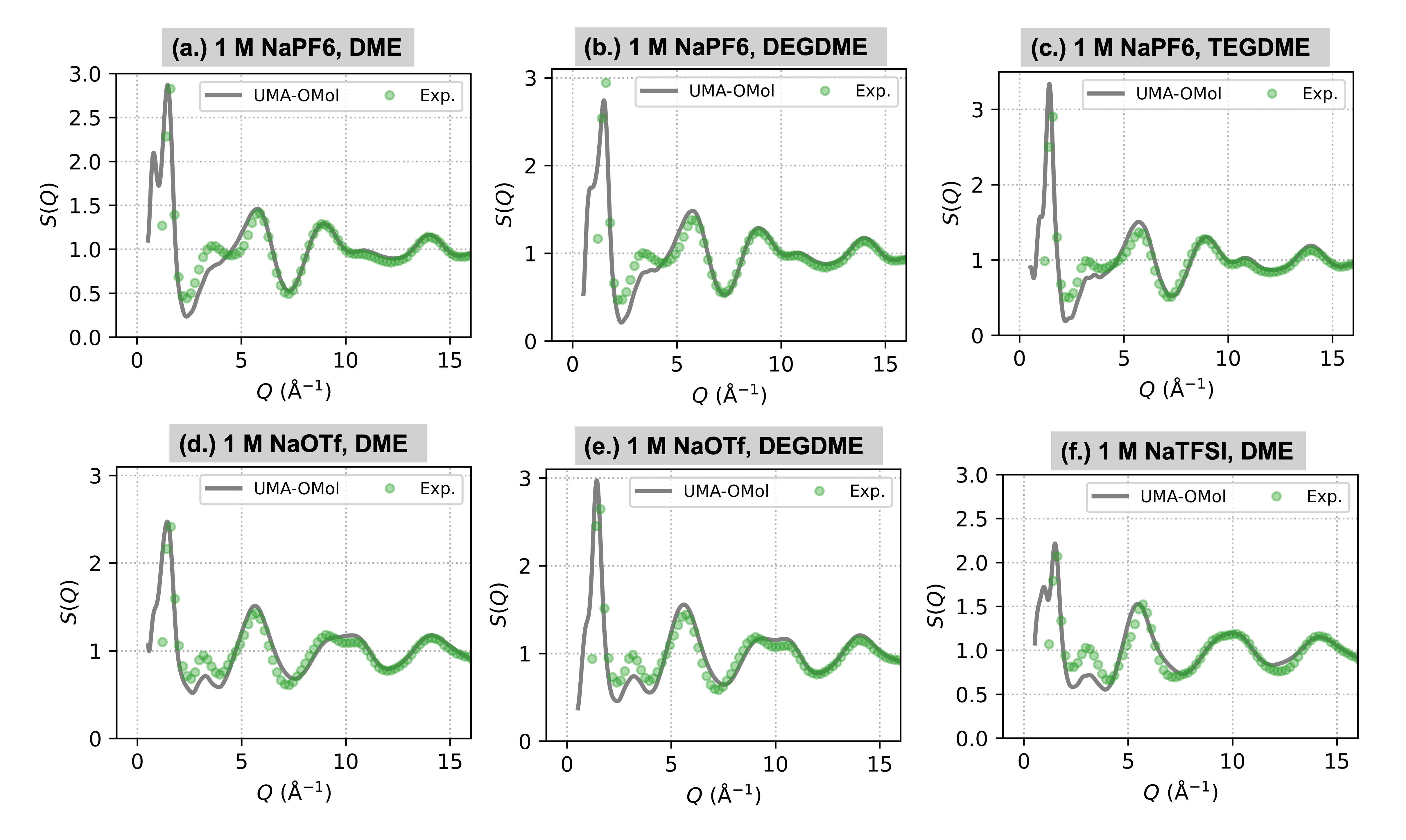}
  \caption{Experimental and UMA-OMol simulated X-ray total scattering structure factors S(Q) for one molar sodium salts glyme electrolytes. Panels compare simulated and experimental S(Q) curves across different anions and solvent chain lengths. }   
  \label{fgr:area}
\end{figure*}

\begin{figure*}[ht!]
\centering
    \captionsetup{name= Figure }
  \includegraphics[height=18.0cm]{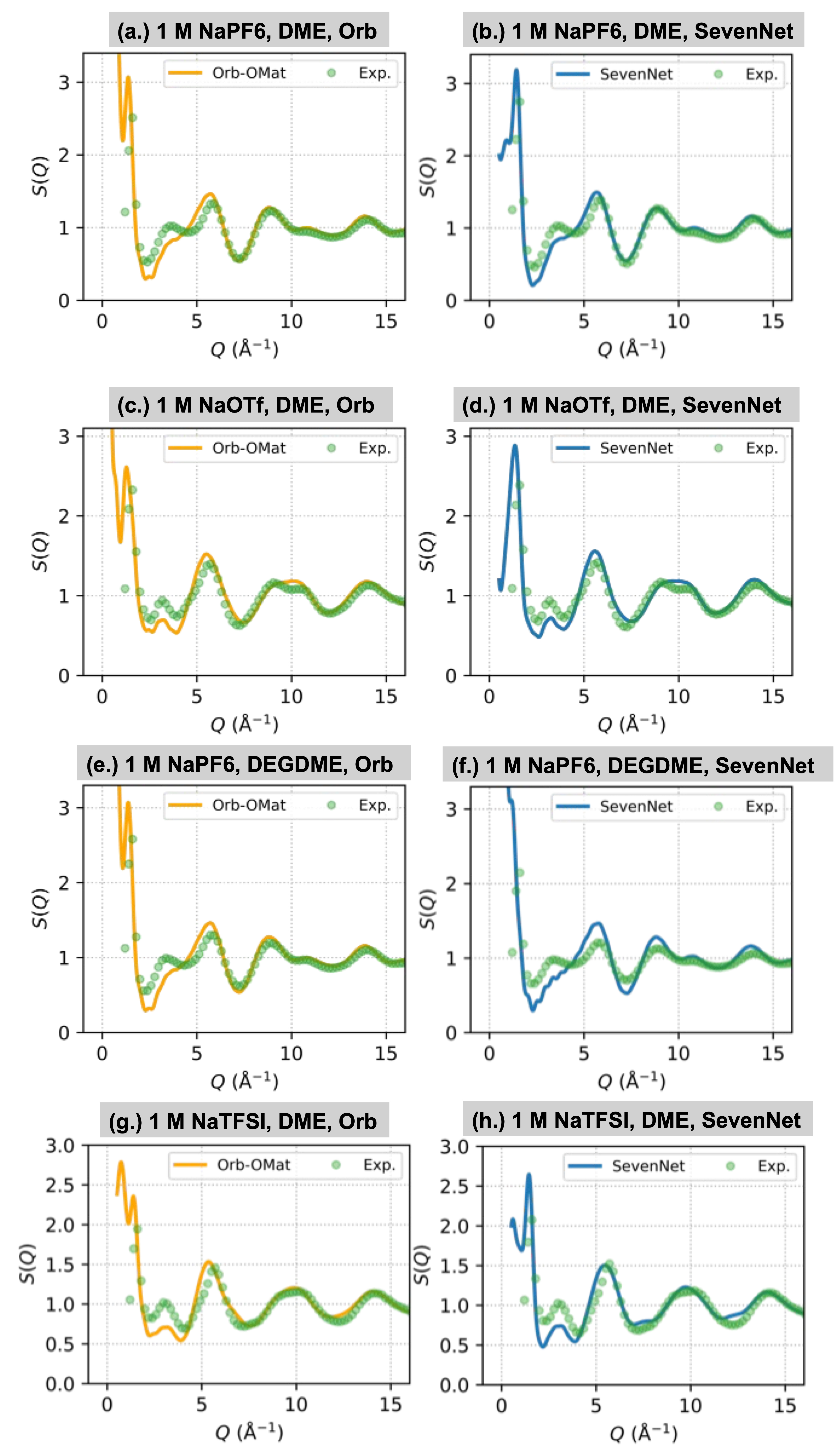}
\caption{Experimental and Orb-OMat or SevenNet simulated X-ray total scattering structure factors S(Q) for one molar sodium salt glyme electrolytes. Panels compare simulated and experimental S(Q) curves across different anions and solvent chain lengths.}
  \label{fgr:area}
\end{figure*}

\begin{figure}[ht!]
\centering
  \includegraphics[height=7.0cm]{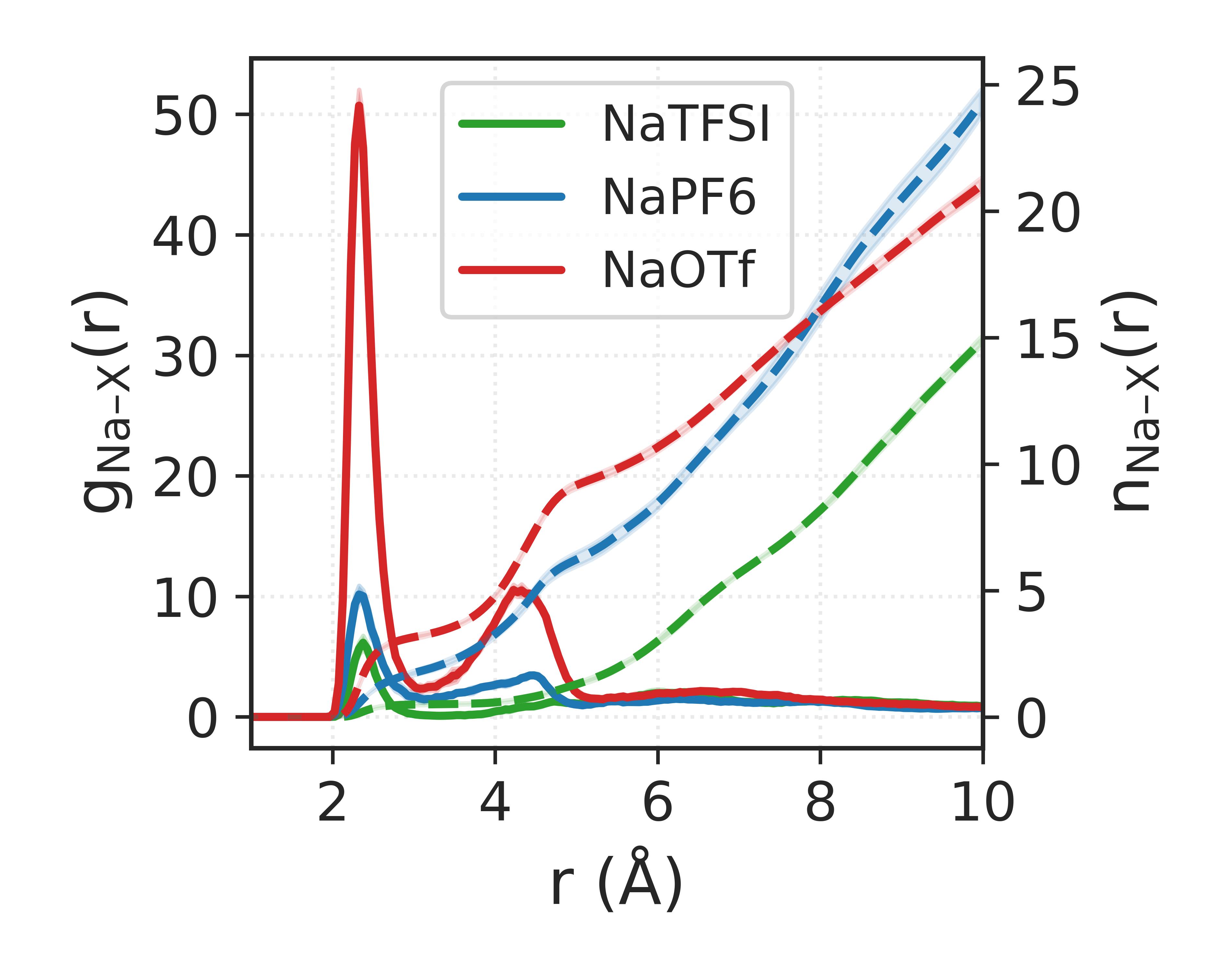}
\caption{Radial distribution functions $g_{\mathrm{Na\text{-}X}}(r)$ and coordination numbers $n_{\mathrm{Na\text{-}X}}(r)$ for Na$^+$ interactions with the dominant coordinating heteroatoms of different anions in DME, namely O in TFSI$^-$, F in PF$_6^-$, and O in OTf$^-$. 
}
\label{fgr:anion}
\end{figure}

\begin{figure}[ht!]
\centering
    \captionsetup{name= Figure }
  \includegraphics[height=5.5cm]{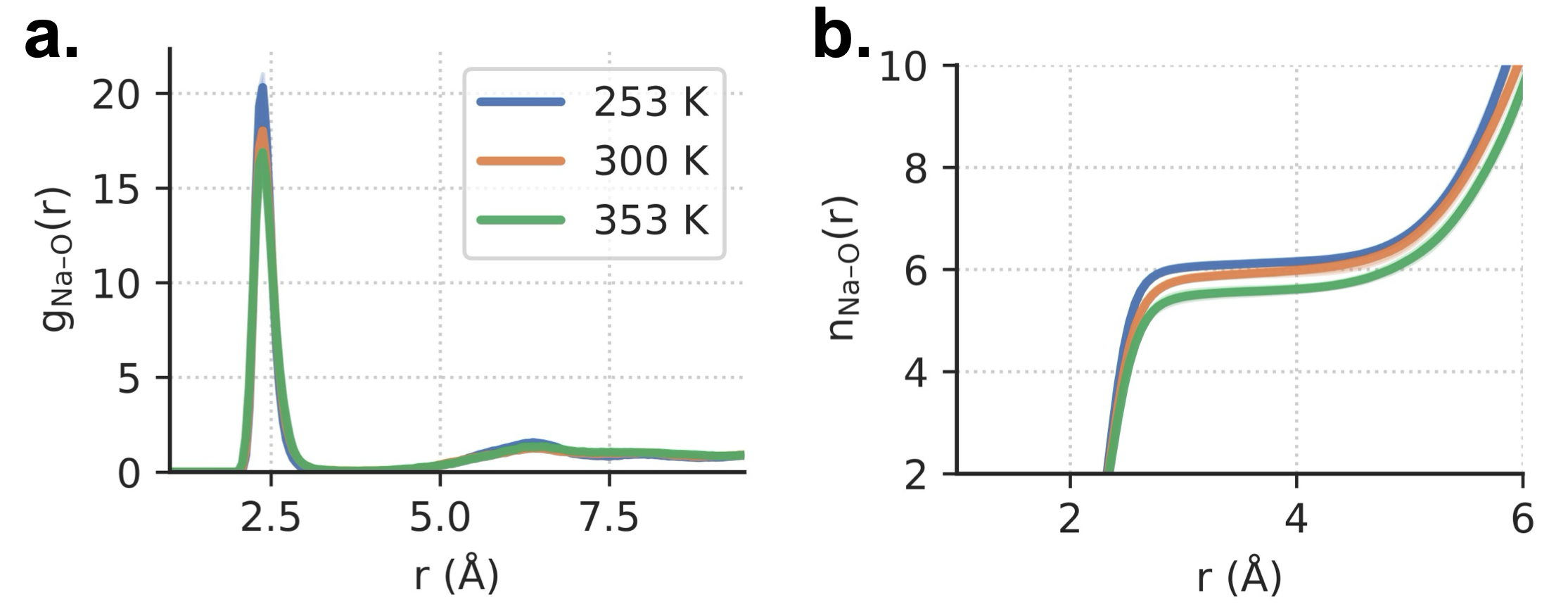}
  \caption{(a) Radial distribution functions for Na–O(solvent) pairs across the simulated temperatures.
(b) Corresponding running coordination numbers derived from the Na–O radial distributions.}  
  \label{fgr:area}
\end{figure}

\begin{figure}[ht!]
\centering
    \captionsetup{name= Figure }
  \includegraphics[height=5.5cm]{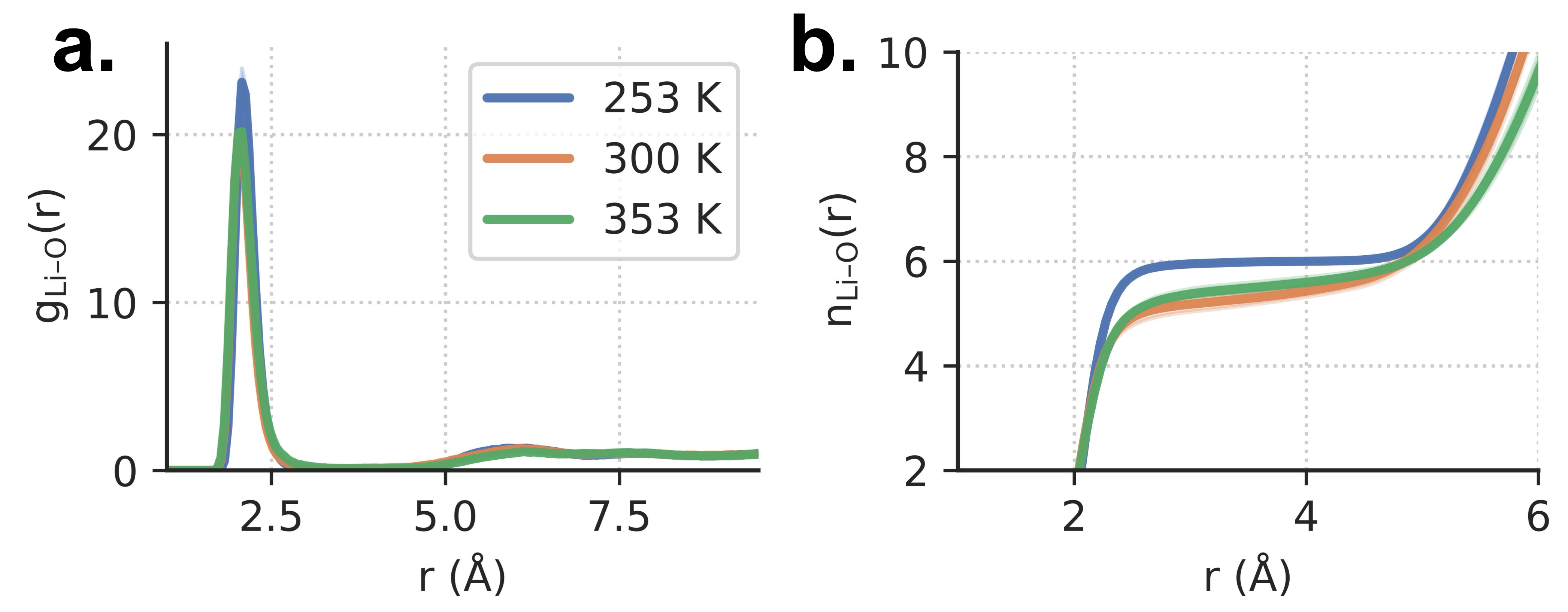}
  \caption{(a) Radial distribution functions for Li–O(solvent) pairs across the simulated temperatures.
(b) Corresponding running coordination numbers derived from the Li–O radial distributions.}  
  \label{fgr:area}
\end{figure}

\begin{figure}[ht!]
\centering
    \captionsetup{name= Figure }
  \includegraphics[height=5cm]{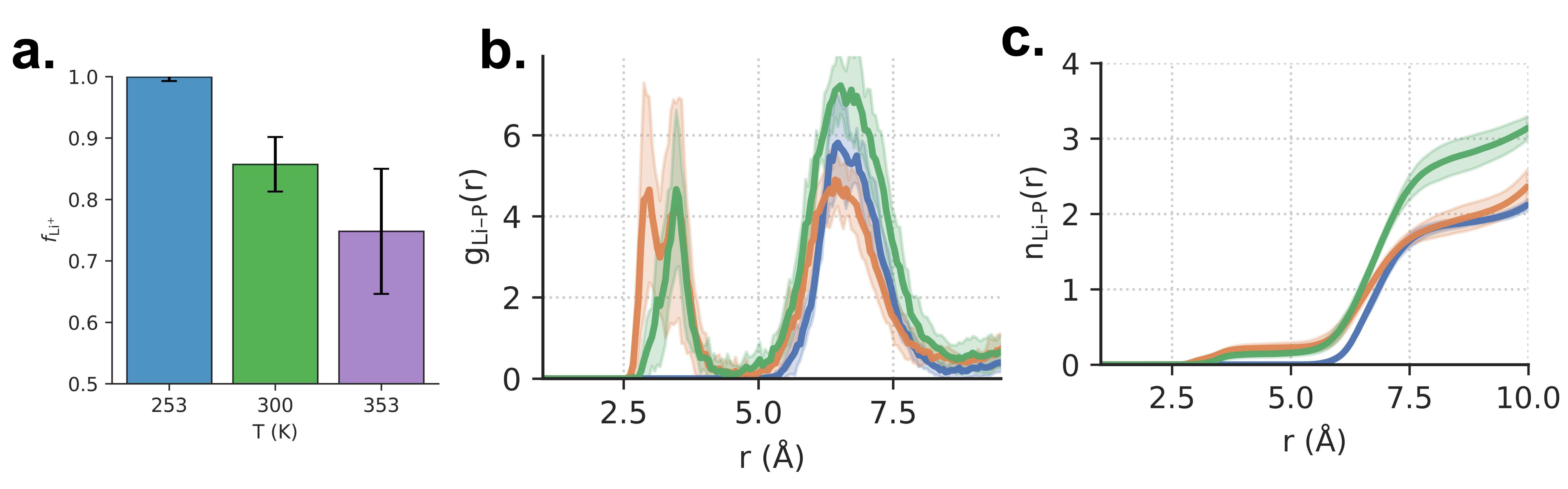}
  \caption{Temperature dependence of \ce{LiPF6} association in DME.
(a) Fraction of free \ce{Li+} ions evaluated for three different simulated temperatures.
(b,c) Li–P radial distribution functions and coordination numbers computed from UMA-OMol simulations. Error estimates are obtained through block averaging. }  
  \label{fgr:area}
\end{figure}

\begin{figure}[ht!]
\centering
    \captionsetup{name= Figure }
  \includegraphics[height=6cm]{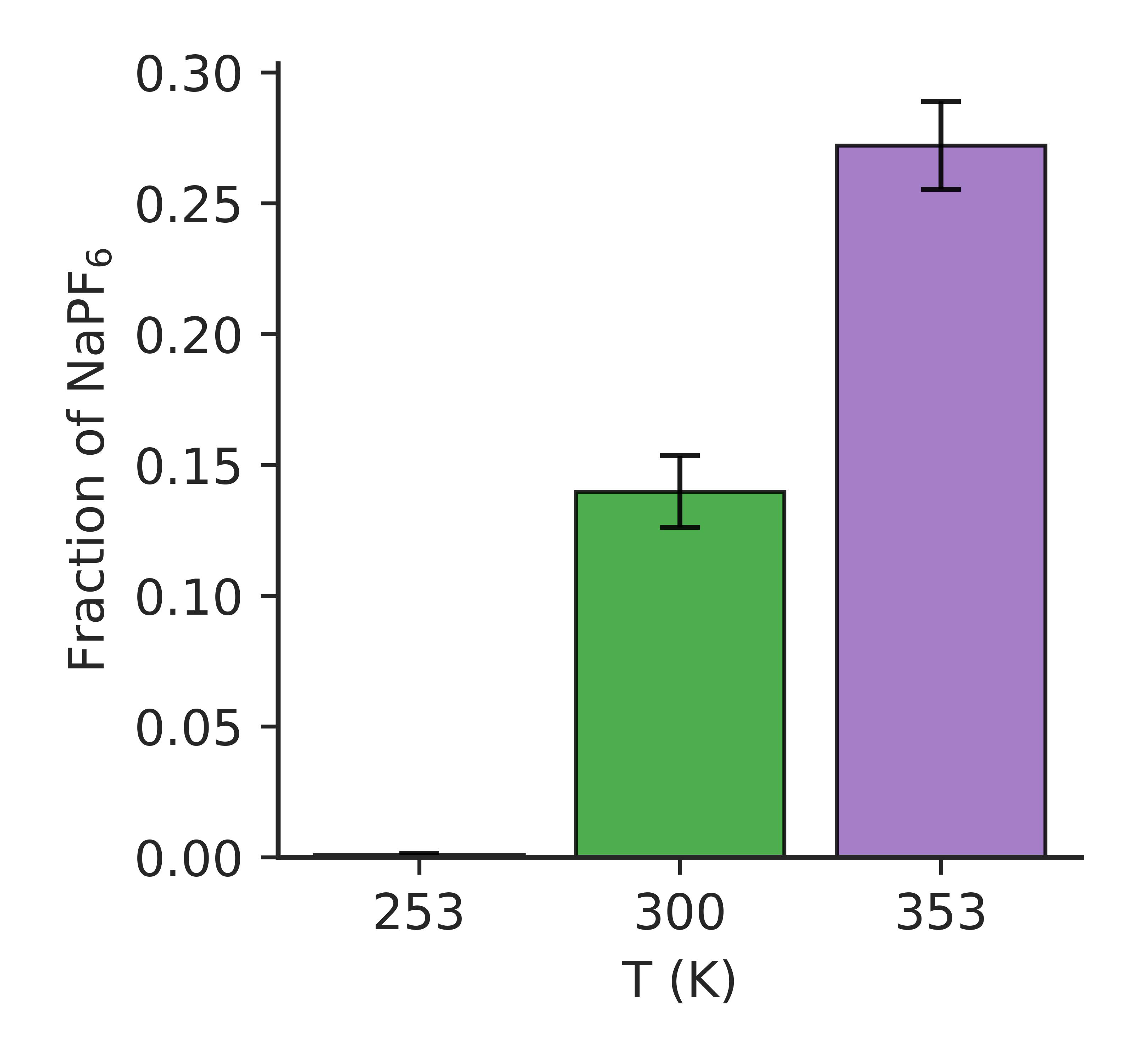}
  \caption{Fraction of \ce{NaPF6} contact ion pairs in 0.5 M \ce{NaPF6} in DME system at different temperatures. }  
  \label{fgr:area}
\end{figure}

\newpage

\begin{figure}[ht!]
\centering
    \captionsetup{name= Figure }
  \includegraphics[height=10cm]{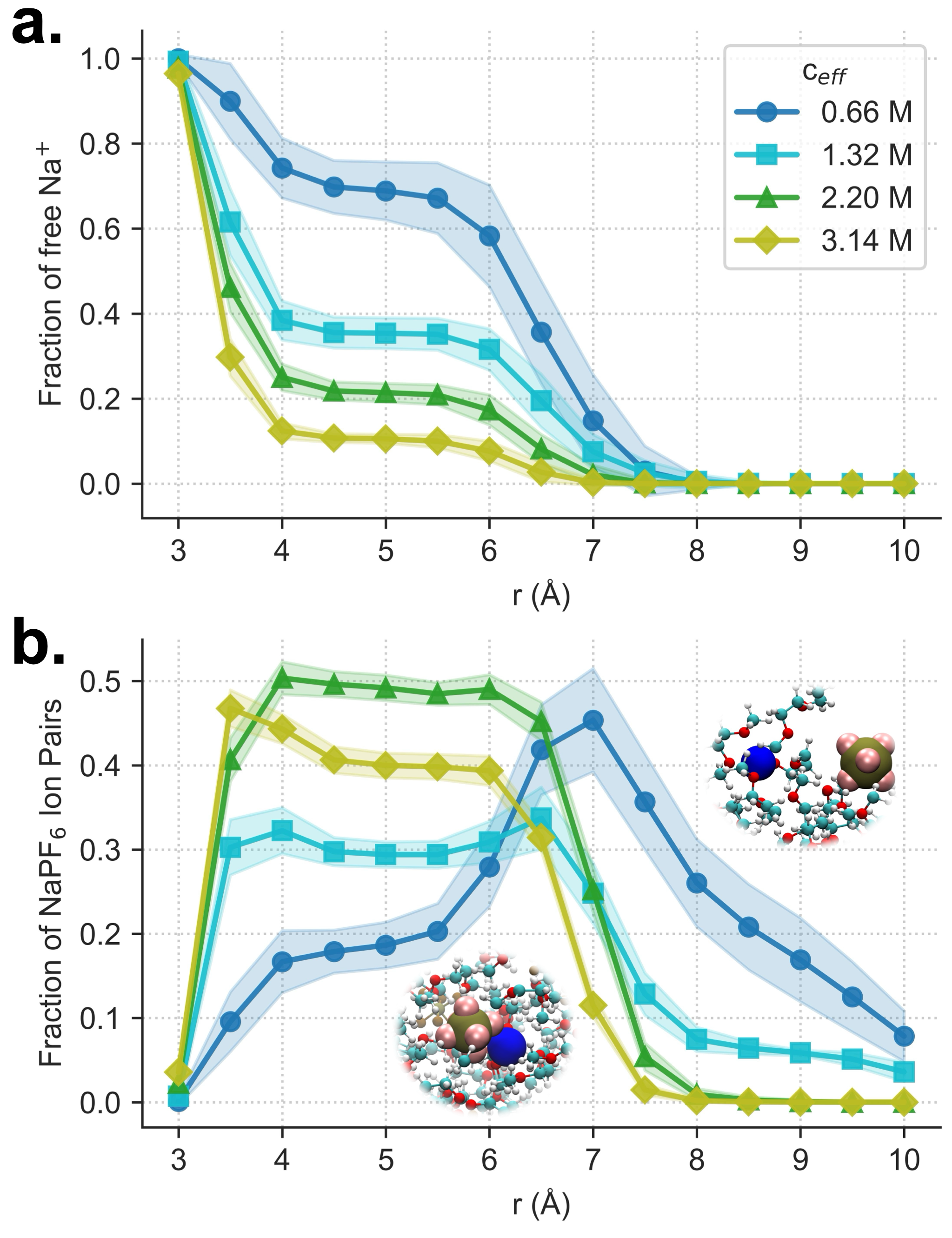}
\caption{Effect of NaPF$_6$ concentration in DME on ion association from UMA-OMol simulations. (a) Fraction of free Na$^+$ ions versus radial cutoff distance at multiple molarities. (b) Fraction of Na$^+$ coordinated to exactly one PF$_6^-$ as a function of cutoff, separating contact ion pairs (CIPs) and solvent-separated ion pairs (SSIPs). Higher concentration shifts the equilibrium toward multi-anion aggregates, reducing free-ion fraction and likely decreasing molar conductivity.}

  \label{fgr:area}
\end{figure}

\begin{figure}[ht!]
\centering
    \captionsetup{name= Figure }
  \includegraphics[height=6cm]{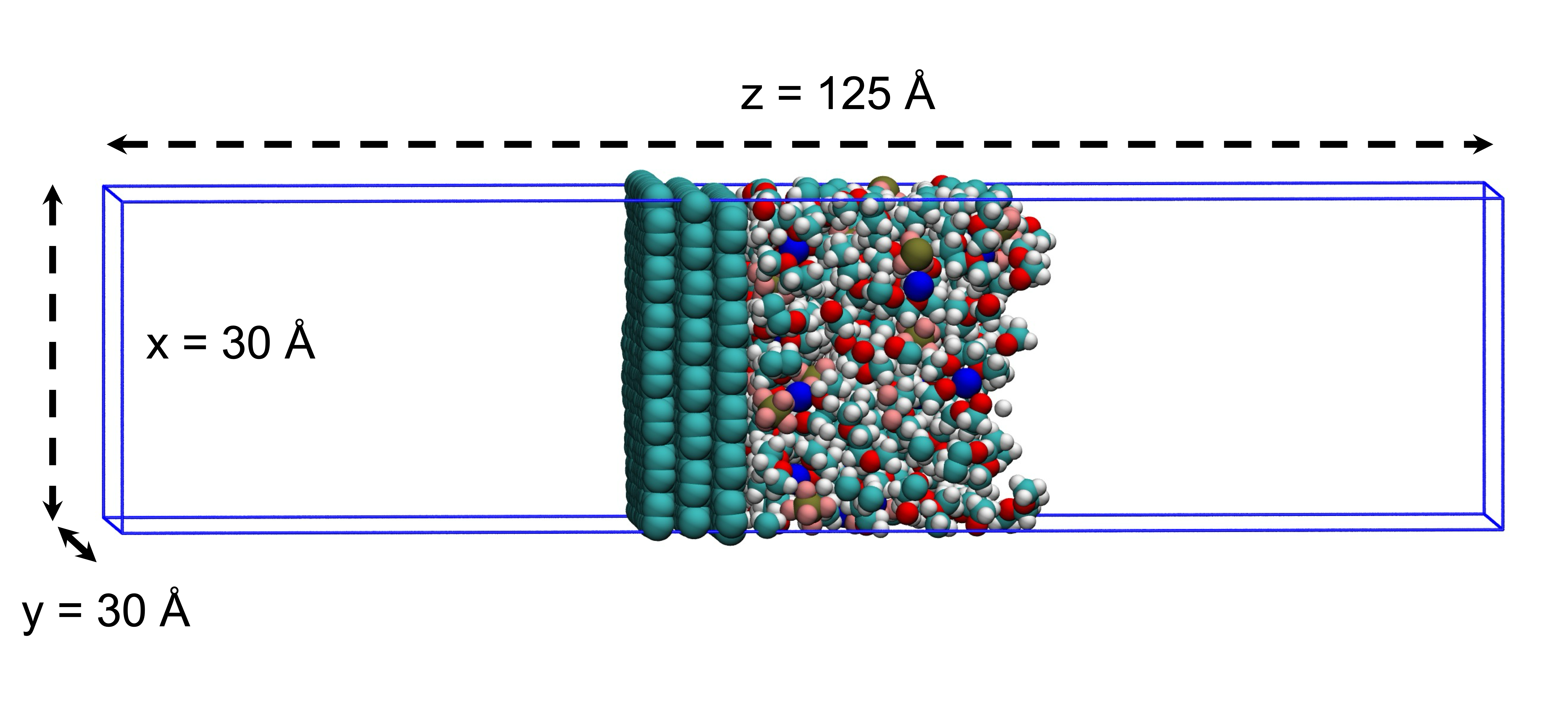}
  \caption{Snapshot of the simulated solid–liquid interfacial system.}
  \label{fgr:area}
\end{figure}

\clearpage
\newpage


\bibliography{achemso-demo}